\documentclass[11pt]{article}
\usepackage{amsmath,fullpage,graphicx,booktabs,palatino,epsfig,fancyhdr,color,amssymb,relsize,setspace}

\usepackage{epstopdf}
\usepackage{url}
\usepackage{bold-extra}
\usepackage{mathtools}
\usepackage{natbib}

\renewcommand{\headrule}{{\hrule width\headwidth height\noheadrule}}

\newcommand{\appsection}[1]{\let\oldthesection\thesection
\renewcommand{\thesection}{Appendix \oldthesection}
\section{#1}\let\thesection\oldthesection}

\begin{document}


\newcommand{\trcvrttl}{VFISV Inversion Code Documentation for {\sc \textbf{solis/vsm}} Pipeline Implementation} 
\newcommand{\trhdrttl}{\solisvsm VFISV Inversion Code Documentation} 
\newcommand{\trauthor}{Brian J. Harker} 
\newcommand{\trnumber}{NSO/NISP-2017-02} 
\newcommand{\trabstct}{Spectral line inversion codes are tools used to interpret spectropolarimetric
                       data; in general, their function is to analyze a set of observed Stokes profiles,
                       and infer the physical properties of the line-formation region in which the Stokes
                       profiles were formed.  For the \solisvsm 6302v pipeline, the inversion is based on
                       a Milne-Eddington model atmosphere, optimized to reproduce the observed Stokes
                       profiles of both Fe {\scshape i} lines at 6301.5\AA\ and 6302.5\AA.  This document
                       provides a detailed overview of the \solisvsm remix of the VFISV (Very Fast Inversion
                       of the Stokes Vector) inversion code, as it is implemented in the \solisvsm pipeline
                       environment.} 

%
\newcommand{\solisvsm}{{\sc solis/vsm} }
\newcommand{\sdohmi}{{\sc sdo/hmi} }
\newcommand{\degree}{$^{\circ}$}
\newcommand{\arcsec}{^{\prime\prime}}
\newcommand{\NoteToSelf}[1]{\textcolor{red}{\texttt{#1}}{}}
\newcommand{\dint}{\displaystyle\int\limits}
\newcommand{\dsum}{\displaystyle\sum\limits}
\newcommand{\deriv}[2]{\frac{\partial{#1}}{\partial{#2}}}
\newcommand{\sderiv}[3]{\frac{\partial{#1}}{\partial{#2}\partial{#3}}}
\newcommand{\onehalf}{\frac{1}{2}}

%
\def\changemargin#1#2{\list{}{\rightmargin#2\leftmargin#1}\item[]}
\let\endchangemargin=\endlist 

%
\setlength\parindent{0pt}

\setcounter{tocdepth}{4}
\setcounter{secnumdepth}{4}

\begin{figure*}[h]
\begin{center}
\vspace{-1.0cm}
\includegraphics[height=4.0cm]{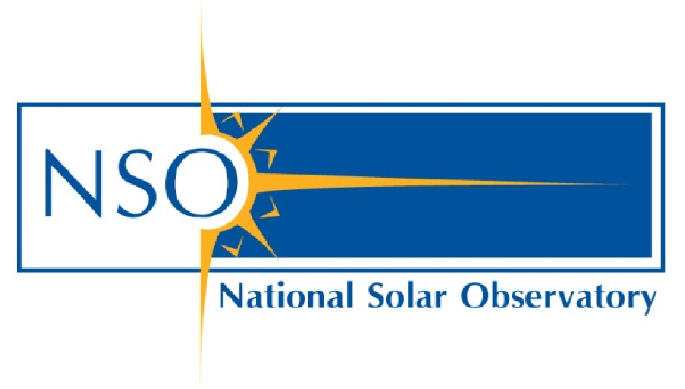}
\end{center}
\end{figure*}
\begin{center}
\vspace{1.0cm}
\begin{LARGE}
\textbf{\trcvrttl}\\
\end{LARGE}
\vspace{1.5cm}
\begin{large}
\trauthor\\
\vspace{0.5cm}
National Solar Observatory\\
\vspace{1.5cm}
\today\\
\vspace{1.5cm}
\hrule
\vspace{0.35cm}
Technical Report No. \textbf{\trnumber}
\vspace{0.25cm}
\hrule
\end{large}
\end{center}

\vfill
\begin{center}
\begin{Large}
\bf{Abstract}
\end{Large}
\end{center}
\begin{quote}
\trabstct
\end{quote}

\clearpage
\renewcommand{\headrule}{{\hrule width\headwidth height\headrulewidth\vskip0.5cm}}
\fancyhead{}
\fancyhead[L]{\trhdrttl}
\fancyhead[R]{\thepage}
\fancyfoot{}
\thispagestyle{fancy}
\vspace*{-0.40cm}

\tableofcontents        


\clearpage
\section{The Milne-Eddington Model Atmosphere}\label{s:me-model}

The polarized radiative transfer equations (PRTE) describe attenuation of the Stokes vector,
$\mathbf{I}_{\lambda}$, of a beam of polarized light as it propagates, in the direction $s$, through
some medium.  Formally, it is given here as

\begin{equation}\label{eq:prte}
\mu\frac{d\mathbf{I}_{\lambda}}{ds} = \mathbf{K}_{\lambda}\left(\mathbf{I}_{\lambda} -
         \mathbf{S}_{\lambda}\right),
\end{equation}

where $\mu$ is the cosine of the heliocentric angle, and $\mathbf{K}_{\lambda}$
and $\mathbf{S}_{\lambda}$ are the propagation matrix and source function vector,
respectively.  Note that, throughout this document, a $\lambda$-subscript denotes a
wavelength-dependent quantity (though it may be dropped for notational clarity).\\

If we model both continuum and line absorption processes, then

\begin{equation}
\mathbf{K}_{\lambda} = \mathbf{1} + \eta_{0}\mathbf{\Phi}_{\lambda} = 
\begin{bmatrix}\eta_{I}&\eta_{Q}&\eta_{U}&\eta_{V}\\
               \eta_{Q}&\eta_{I}&\rho_{V}&-\rho_{U}\\
               \eta_{U}&-\rho_{V}&\eta_{I}&\rho_{Q}\\
               \eta_{V}&\rho_{U}&-\rho_{Q}&\eta_{I}\end{bmatrix}_{\lambda}
\end{equation}

where $\mathbf{1}$ is a $4 \times 4$ identity matrix, $\eta_{0}$ is the ratio of line-to-continuum
absorption coefficients, and the matrix $\mathbf{\Phi}_{\lambda}$ includes both absorption and
magneto-optical effects, parameterized by the magnetic and thermodynamic properties of the
model atmosphere under the classical Zeeman effect regime.  Expressions for these matrix
elements may be found in \citet{deglinnocenti:2004} and references therein.  They are reproduced
here for completeness, with the wavelength dependence implied but dropped from the notation for
clarity.

\begin{align}
\eta_{I} &= 1 + \onehalf\eta_{0}\left[\phi_{\pi}\sin^{2}\gamma + \onehalf\left(\phi_{\sigma_{r}} + \phi_{\sigma_{b}}\right)\left(1 + \cos^{2}\gamma\right)\right]\\
\eta_{Q} &= \onehalf\eta_{0}\left[\phi_{\pi} - \onehalf\left(\phi_{\sigma_{b}} + \phi_{\sigma_{r}}\right)\right]\sin^{2}\gamma\cos2\chi\\
\eta_{U} &= \onehalf\eta_{0}\left[\phi_{\pi} - \onehalf\left(\phi_{\sigma_{b}} + \phi_{\sigma_{r}}\right)\right]\sin^{2}\gamma\sin2\chi\\
\eta_{V} &= -\onehalf\eta_{0}\left(\phi_{\sigma_{r}} - \phi_{\sigma_{b}}\right)\cos\gamma\\
\rho_{Q} &= \onehalf\eta_{0}\left[\psi_{\pi} - \onehalf\left(\psi_{\sigma_{b}} + \psi_{\sigma_{r}}\right)\right]\sin^{2}\gamma\cos2\chi\\
\rho_{U} &= \onehalf\eta_{0}\left[\psi_{\pi} - \onehalf\left(\psi_{\sigma_{b}} + \psi_{\sigma_{r}}\right)\right]\sin^{2}\gamma\sin2\chi\\
\rho_{V} &= -\onehalf\eta_{0}\left(\psi_{\sigma_{r}} - \psi_{\sigma_{b}}\right)\cos\gamma
\end{align}

The $\phi$ and $\psi$ terms are the absorption and anomalous dispersion profiles (and are functions of wavelength) for the 
$\pi$, $\sigma_{r}$, and $\sigma_{b}$ Zeeman components involved in the transition which gives rise to the absorption
line itself.  As can be seen above, the matrix elements themselves depend on the vector magnetic field explicitly through
the inclination, $\gamma$, and azimuthal angle, $\chi$, and also have a dependence on the field strength through the Voigt
and Faraday-Voigt spectral line shapes, given by:

\begin{align}
H_{\lambda}(a_{\mathrm{dc}},v) &= \frac{a_{\mathrm{dc}}}{\pi}\dint_{-\infty}^{+\infty}
        \frac{e^{-t^{2}}}{a_{\mathrm{dc}}^{2} + (v-t)^{2}}\mathrm{d}t\\
F_{\lambda}(a_{\mathrm{dc}},v) &= \frac{1}{\pi}\dint_{-\infty}^{+\infty}
        \frac{(v-t)e^{-t^{2}}}{a_{\mathrm{dc}}^{2} + (v-t)^{2}}\mathrm{d}t\\
v &= \frac{\lambda - \lambda_{0}(1+v_{\mathrm{los}}/c) \pm \Delta\lambda_{Z}}{\Delta\lambda_{D}},\label{eq:v}
\end{align}

where $\lambda_{0}$ is the line-center wavelength, $v_{\mathrm{los}}$ is the relative line-of-sight (LOS)
velocity between source and observer, $a_{\mathrm{dc}}$ is the atomic damping constant of the line,
$\Delta\lambda_{D}$ is the Doppler line-width, and $\Delta\lambda_{Z}$ is the Zeeman splitting of the line
(which is itself a function of the atomic energy levels of the transition as well as magnetic field strength,
$B$).  The absorption ($\phi$) and anomalous dispersion profiles ($\psi$) are then calculated as:

\begin{align}
\phi_{i} &= \sum_{j=1}^{N_{z}}S_{j}^{(i)}H_{\lambda}(a_{\mathrm{dc}},v_{j})\\
\psi_{i} &= \sum_{j=1}^{N_{z}}S_{j}^{(i)}F_{\lambda}(a_{\mathrm{dc}},v_{j}),
\end{align}

where $i = \lbrace\pi,\sigma_{r},\sigma_{b}\rbrace$, $N_{z}$ is the number of Zeeman subcomponents
of the $i^{th}$ component, and $S_{j}^{(i)}$ is the Zeeman strength of the $j^{th}$ Zeeman subcomponent.
The strengths are functions of the quantum numbers of the upper and lower energy levels of the Zeeman
subcomponents involved in the transition.  Analytical expressions for $S_{j}^{(i)}$ are given in
Table 3.1 of \citet{deglinnocenti:2004} for all allowed transitions.  For a normal Zeeman triplet (like
the Fe {\sc i} 6302.5\AA\ line), there is only one $\pi$, $\sigma_{r}$, and $\sigma_{b}$ Zeeman component,
so $S_{j}^{(i)} = 1\ \forall\ i,j$.  However, for an anomalous Zeeman triplet (like the Fe {\sc i}
6301.5\AA\ line), there are multiple Zeeman subcomponents, such that we require:

\begin{equation}
\sum_{j=1}^{N_{z}}S_{j}^{(i)} = 1,\ \forall\ i.
\end{equation}

The source function vector ($\mathbf{S}_{\lambda}$ in Equation (\ref{eq:prte})) describes the ratio of
emission to absorption in the beam, and includes both continuum and line contributions, so that

\begin{equation}
\mathbf{S}_{\lambda} = S_{c}\hat{\mathbf{e}} + \eta_{0}S_{l}\mathbf{\Phi}_{\lambda}\hat{\mathbf{e}},
\end{equation}

where $\hat{\mathbf{e}} = \left(1,0,0,0\right)^{T}$, $S_{c}$ is the continuum source
function, and $S_{l}$ is the line source function.  Assuming local thermodynamic equilibrium
reduces both the continuum and line source functions to the Planck function at the local temperature,
$B_{\lambda}(T)$.  Adopting a Milne-Eddington (ME) relation for the source function variation
as a linear function of optical depth,

\begin{equation}
S_{c} = S_{l} = B_{\lambda}(T) = S_{0} + S_{1}\tau = S_{0}(1 + \beta_{0}\tau),
\end{equation}

the PRTE admits an analytical solution for the model Stokes profiles,
$\mathbf{I}_{\lambda}^{\mathrm{M}}$, given here as

\begin{align}
\frac{\mathbf{I}_{\lambda}^{\mathrm{M}}}{I_{c}} &= \left[ \left(1-\beta\right)\mathbf{1} +
        \beta\left( \mathbf{1} + \eta_{0}\mathbf{\Phi}_{\lambda} \right)^{-1}
        \right]\hat{\mathbf{e}}\\
\beta &= \frac{\mu\beta_{0}}{1+\mu\beta_{0}}.
\end{align}

Note that $\beta_{0}=S_{1}/S_{0}$ represents the inverse of the characteristic
length scale over which the source function changes appreciably, and $I_{c}$ denotes the
observed local continuum intensity.  Note also that the Eddington-Barbier
approximation states that the emergent continuum intensity at a position $\mu$ on the solar disc
is given by $S_{c}(\tau=\mu)$, so that we have the relation $I_{c} = S_{0} + \mu S_{1}$, coupling
the (observed) continuum intensity with the source function coefficients.  The source function
coefficients \textit{must} obey this constraint at all points in the inversion.\\

At a modest 1 arcsec spatial sampling, \solisvsm does not fully resolve magnetic structures.
Therefore, we introduce a geometrical factor which represents the fraction of the pixel occupied
by magnetic field, and as such, must be in the range $[0,1]$.  This is the magnetic filling-factor,
$\alpha$, and is a zeroth-order attempt at accounting for limited spatial resolution, by allowing
the pixel to contain a mixture of Stokes profiles from both magnetic and non-magnetic atmospheres
coexisiting within the same pixel.  The Stokes vector then becomes a linear superposition of
the magnetic and non-magnetic profiles, weighted by $\alpha$,

\begin{equation}
\mathbf{I}_{\lambda}^{\mathrm{M}} \leftarrow \alpha\mathbf{I}^{\mathrm{M}}_{\lambda} +
        (1-\alpha)I^{\mathrm{nm}}_{\lambda}\hat{\mathbf{e}}.
\end{equation}

The non-magnetic profile, $I^{\mathrm{nm}}_{\lambda}$, is synthesized from the same
model parameter vector as the magnetic profile, but magnetic field strength is 
set to zero.  This is preferable to using (e.g.) an average observed quiet-sun profile
constructed from low-polarization Stokes $I$ profiles in the surroundings, since we would
have to deconvolve the instrumental profile from the average quiet-sun profile before mixing
it with the magnetic profile, and then re-convolve with the instrumental profile.  We like to avoid
deconvolution at all costs, thank you very much.  To account for the smearing effects of the 
\solisvsm instrumental profile, the synthetic Stokes vector $\mathbf{I}^{\mathrm{M}}_{\lambda}$ is
convolved with a Gaussian kernel with a half-width at half-maximum (HWHM) of 22.5m\AA.  This is 
equivalent to considering a macroturbulent velocity, $v_{\mathrm{mac}}$, of $1.3081\ \times\ 10^{5}\ \mathrm{cm/sec}$,
which is held constant during the inversion.\\

As shown above, the Milne-Eddington (ME) atmosphere is characterized by the following
assumptions: 1) the line-formation region is under local thermodynamic equilibrium,
2) the source function varies linearly with optical depth, 3) all other physical
properties are constant over the line-formation region, and 4) the polarized radiative
transfer is described by the classical Zeeman effect.  As such, it is a simple
(but widely-used) model.  Since we do not consider gradients with respect to optical depth
of any parameter except the source function, the ME atmosphere represents a kind of average
of the true parameters over the height of line-formation.  Under these assumptions, the PRTE
admits an analytical solution, known as the Unno-Rachkovsky solutions
\citep{unno:1956,rachkovsky:1962,rachkovsky:1963,rachkovsky:1967},
which are functions parameterized by the free model parameter vector,
$\mathbf{p} = [B,\gamma,\chi,v_{\mathrm{los}},\Delta\lambda_{D},a_{\mathrm{dc}},\eta_{0},S_{0},\mu S_{1},\alpha]$.
The model parameters are listed and summarized below for clarity:

\begin{itemize}
  \renewcommand\labelitemi{\scriptsize$\blacksquare$}
  \item $B$: Magnetic field strength
  \item $\gamma$: Inclination of the magnetic field vector relative to the LOS
  \item $\chi$: Azimuthal angle of the transverse component of the magnetic field vector
  \item $v_{\mathrm{los}}$: Relative LOS velocity between source and observer
  \item $\Delta\lambda_{D}$: Doppler line width
  \item $a_{\mathrm{dc}}$: Atomic damping constant of the line
  \item $\eta_{0}$: Line-to-continuum opacity ratio
  \item $S_{0}$: Source function continuum
  \item $\mu S_{1}$: heliocentric $\mu$ $\times$ Source function gradient
  \item $\alpha$: Magnetic filling-factor
  \item $v_{\mathrm{mac}}$: Macroturbulent velocity
\end{itemize}

\subsection{Levenberg-Marquardt Inversion}\label{s:lm}

The VFISV (Very Fast Inversion of the Stokes Vector) inversion code executes a general
non-linear least-squares optimization of a Milne-Eddington model atmosphere to reproduce
Stokes profiles which best fit the actual observations.  It has been tuned for speed
to enable the \solisvsm 6302v pipeline to invert \textit{every} pixel on the solar
disc in a short time ($\sim$ 10 minutes).\\

The function to be minimized by the inversion algorithm is the following
$\chi^{2}$-like merit function, which quantifies the agreement between the Stokes vector
($\mathbf{I}^{M}$) synthesized in the model atmosphere (parameterized by $\mathbf{p}$)
and the observations ($\mathbf{I}^{O}$), given here explicitly as

\begin{equation}\label{eqn:chi2}
\chi^{2}(\mathbf{p}) = \frac{1}{\nu}\dsum_{i} \dsum_{j=1}^{N_{\lambda}}
        w^{2}_{ij}\left[\mathbf{I}^{\mathrm{O}}_{i}(\lambda_{j}) -
        \mathbf{I}^{\mathrm{M}}_{i}(\lambda_{j};\mathbf{p})\right]^{2},
\end{equation}

where $i=I,Q,U,V$. The number of degrees of freedom in the optimization is given by $\nu$, and
$N_{\lambda}$ is the number of observed wavelengths spanning the bandpass containing the lines
to be inverted.  The quantities $w_{ij}$ are weighting factors (entering quadratically),
traditionally used to adjust the contribution of different wavelengths to the total deviation
between the modelled and observed spectral line, and are discussed further in Section
\ref{ss:weights}.\\

Given an initial guess for the model parameters, the inversion code uses a Levenberg-Marquardt
non-linear least squares optimization algorithm to perform the inversion.  At each step,
perturbations to the current ME model parameters are calculated, and the new set of model
parameters are used to synthesize Stokes profiles.  The $\chi^{2}$ merit function is 
calculated via Equation (\ref{eqn:chi2}) and tested against the current value.  If an
improvement (reduction) is found, the perturbations are accepted and the model parameters
are updated for the next iteration.  If not, new perturbations are calculated.  This iterative
improvement proceeds until one or more convergence criteria are satisfied (more on this in Section
\ref{ss:conv}), at which time the algorithm outputs the current model parameters as the solution
to the inversion problem; i.e., the physical parameters of the ME model atmosphere in which the
observed Stokes profiles were formed.

\clearpage
\section{Implementation}

The VFISV code used in the \solisvsm 6302v production pipeline is meant to handle standard
\solisvsm Level-0.5 spectra contained in FITS files.  For full-disc observations, this 
consists of 2048 fully-calibrated (dark-corrected, flat-fielded, polarization-calibrated)
scanlines, each containing 128 wavelength samples for 4 Stokes profiles for 2048 spatial pixels
along the spectrograph slit.  This Section focuses on various issues related to the
implementation of the code (libraries, compilation) in the 6302v pipeline environment.\\

\subsection{Compiling the VFISV Code}

This Section contains a short list of implementation details relating to what is required
to compile the VFISV code.

\begin{itemize}
  \renewcommand\labelitemi{\scriptsize$\blacksquare$}
  \item Language: Fortran 90 (with a single utility function written in C).
  \item Architecture: Developed for 64-bit machines.
  \item MPI parallelization: VFISV depends on the Message Passing Interface (MPI) for parallel
                             processing of individual scanlines.  The MPI interface
                             is provided by MPICH-2, the MPI implementation from Argonne National
                             Laboratory.  MPICH-2 is a high performance and widely portable implementation
                             of the MPI standard.
  \item Compiler: On the machine used for development of the code, MPICH-2 for Fortran was built using the
                  Intel \texttt{ifort} compiler.  As such, the MPI wrapper compiler \texttt{mpif90}
                  has a backend based on \texttt{ifort}.  No (serious) attempt was made at compiling/running
                  the code with any other compiler.  Fair warning.  The single utility function 
                  written in C is compiled with the GNU \texttt{gcc} compiler.
  \item Compiler Flags/Optimizations: 
        \begin{itemize}
        \renewcommand\labelitemii{\textbullet}
          \item \texttt{-fast} : Maximizes speed across the entire program.  On Itanium(R)-based
                                 systems, this option sets options -ipo, -O3, and  -static.   On
                                 IA-32  and  Intel(R)  EM64T  systems,  this option sets options
                                 -ipo, -O3, -no-prec-div, -static, and -xP.  Note that programs
                                 compiled with the  -xP  option  will  detect non-compatible 
                                 processors and generate an error message during execution.  This 
                                 optimization  has a dramatic impact on runtime; the code was tested
                                 without this optimization flag, and the results were identical to
                                 those obtained with this optimization, although they were obtained
                                 \textit{much} slower.
          \item \texttt{-prec-div} : Improves precision of floating-point divides; it has some speed
                                     impact.  With some optimizations, such as -xN and -xB, the
                                     compiler may change floating-point division computations into
                                     multiplication by the reciprocal of the denominator. For example,
                                     A/B is computed as A * (1/B) to improve the speed of the computation.
                                     However, sometimes the value produced by this transformation is
                                     not as accurate as full IEEE division. When it is important  to
                                     have fully precise IEEE division, use -prec-div to disable the
                                     floating-point division-to-multiplication  optimization.   The
                                     result is more accurate, with some loss of performance.
          \item \texttt{-check all} : Enables all \texttt{-check} options.  NOTE: while not routinely compiled
                                      with this flag, it can be helpful to occasionally check for
                                      certain conditions at runtime (i.e. after major changes to the 
                                      code), like out-of-bounds array subscripts.
          \item \texttt{-warn all} : Enables all diagnostic warning messages.  NOTE: while not routinely
                                     compiled with this flag, it can be helpful to occasionally check for
                                     certain conditions at compile time (i.e. after major changes to the 
                                     code), like unused variables, etc.
          \item \texttt{-g} : Enables debugging support.  Not routinely compiled with this flag.
          \item \texttt{-p} : Enables \texttt{gprof} profiling support.  Not routinely compiled with this flag.
        \end{itemize}
  \item Required libraries: MPI, LAPACK, CFITSIO
  \item Makefile: The Makefile included with the code is sufficiently general that it should work for
                  most machines/environments, provided the right paths are set.  To run in other environments,
                  one just needs to make sure that the required libraries (MPI, LAPACK, CFITSIO) are installed,
                  and the path(s) to these libraries are correctly specified.
        \begin{itemize}
        \renewcommand\labelitemii{\textbullet}
          \item To compile the code, simply run \texttt{make vfisv}.
          \item To install the \texttt{vfisv.x} executable into the SOLIS \texttt{bin}
                directory on a production/reprocessing machine, simply run \texttt{make install}.  Note that
                this strips the ``\texttt{.x}'' file extension to produce the \texttt{vfisv} binary executable
                in the \texttt{bin} directory.
          \item To remove all compiled files, modules, objects (etc.), simply run \texttt{make clean}.
          \item To create a datestamped tarball of the VFISV codebase and support files (IDL functions, e.g.), simply
                run \texttt{make backup}.
        \end{itemize}
\end{itemize}

\clearpage
\section{Running the VFISV Code}

This Section contains information detailing how to run the VFISV code.  Specifically, I present
here summaries of the calling syntax, with appropriate command line arguments, and give
examples of how to run the code in various modes.

\subsection{Calling Syntax}
The default syntax for running the VFISV inversion code is as follows (note that square brackets
denote optional arguments):\\
\\
\texttt{\$ mpirun -np NP /home/solis/bin/vfisv [-v] [-vv] [-P] -i /path/to/Level-0.5/dir\\
-w /path/to/Level-1/output/tmp/dir [-s s1 [s2]] [-p p1 [p2]]}\\
\\
where \texttt{NP} is the number of MPI processes requested for the inversion.  Note that the supervisor MPI rank
does no work inverting scanlines itself, so only \texttt{NP-1} scanlines can be inverted simultaneously (see
Section \ref{ss:m-s}).

\subsubsection{Command Line Arguments}
\begin{itemize}
  \renewcommand\labelitemi{\scriptsize$\blacksquare$}
  \item \texttt{-i} : [REQUIRED] Specify the path to the directory containing \solisvsm Level-0.5 scanline FITS files.
  \item \texttt{-w} : [REQUIRED] Specify the path to the directory which will contain \solisvsm Level-1 FITS file output
                      from the inversion.  This directory will be created if it does not already exist.
  \item \texttt{-s} : [OPTIONAL] Specify a range of scanlines to invert.  This is useful for inverting only a fraction
                      of a full-disc observation, or for inverting entire area-scan observations.  If only a single scanline
                      index is provided (\texttt{-s s1}), the code will invert only that scanline; if two indices are provided
                      (\texttt{-s s1 s2}), the code will invert all scanlines between \texttt{s1} and \texttt{s2}, inclusive.
  \item \texttt{-p} : [OPTIONAL] Specify a range of pixels to invert.  This is useful (in combination with \texttt{-s})
                      for inverting only a rectangular subset of the available dataset (e.g., if you're 
                      interested in a particular active region).  If only a single pixel index is provided (\texttt{-p p1}),
                      the code will invert only that pixel; if two indices are provided (\texttt{-p p1 p2}), the code will
                      invert all pixels between \texttt{p1} and \texttt{p2}, inclusive.
  \item \texttt{-v} : [OPTIONAL] Switch on verbose messages (for debugging).  NOTE: this will produce \textit{a lot} of
                      screen output, so it is not suitable for regular full-disc inversions, but rather
                      intended to be used on a single-pixel basis for development and/or diagnosing potential
                      issues.
  \item \texttt{-vv}: [OPTIONAL] Switch on extra verbose messages (for debugging).  This produces much more screen
                      output than the \texttt{-v} option.  \texttt{-vv} implies \texttt{-v}.
  \item \texttt{-P} : [OPTIONAL] Switch on ``preview mode'', which rapidly generates a Level-1 quicklook file from
                      non-defringed Level-0.5 spectra.  This is \textit{only} used by the observers to quickly get
                      feedback on instrument pointing, (e.g.) during area-scans.
\end{itemize}

\subsubsection{Examples}
A full-disc 6302v observation was taken on 08 July 2015.  To invert the Level-0.5 spectra for this observation
(contained in the directory \texttt{k4v9s150708t175025\_oid114363777241750\_cleaned}) using 16 MPI processes, and
place the resulting Level-1 file(s) in the appropriate tmp directory, run:\\

\texttt{\$ mpirun -np 16 /home/solis/bin/vfisv -i \\
/path/to/v9s/201507/k4v9s150708/k4v9s150708t175025\_oid114363777241750\_cleaned\\
-w /path/to/vsm/tmp/k4v91150708t175025}\\

To invert only those scanlines between 512 and 1536 (inclusive), run:\\
\texttt{\$ mpirun -np 16 /home/solis/bin/vfisv -i \\
/path/to/v9s/201507/k4v9s150708/k4v9s150708t175025\_oid114363777241750\_cleaned\\
-w /path/to/vsm/tmp/k4v91150708t175025 -s 512 1536}\\

To invert only those scanlines between 512 and 1536, and only in Camera A (e.g.), run:\\
\texttt{\$ mpirun -np 16 /home/solis/bin/vfisv -i \\
/path/to/v9s/201507/k4v9s150708/k4v9s150708t175025\_oid114363777241750\_cleaned\\
-w /path/to/vsm/tmp/k4v91150708t175025 -s 512 1536 -p 1 1024}\\

To invert only those scanlines between 512 and 1536, and only in Camera B (e.g.), run:\\
\texttt{\$ mpirun -np 16 /home/solis/bin/vfisv -i \\
/path/to/v9s/201507/k4v9s150708/k4v9s150708t175025\_oid114363777241750\_cleaned\\
-w /path/to/vsm/tmp/k4v91150708t175025 -s 512 1536 -p 1024 2048}\\

To invert only a single pixel with verbose messages for debugging, run:\\
\texttt{\$ mpirun -np 2 /home/solis/bin/vfisv -i \\
/path/to/v9s/201507/k4v9s150708/k4v9s150708t175025\_oid114363777241750\_cleaned\\
-w /path/to/vsm/tmp/k4v91150708t175025 -s 512 -p 768 -v}\\

NOTE: for single-pixel inversion, you \textit{must} use \texttt{mpirun -np 2}, or
you will get an error message complaining that too many MPI processes would be used:\\

\texttt{\indent \$ [ERROR] MPI NUM\_RANKS $>$ Number of scanlines to invert!"\\
        \indent \$ [ERROR] Check your mpirun command line..."}\\

In general, you must use an appropriate number of MPI ranks for the inversion; if the number of
MPI ranks is undersubscribed relative to the number of scanlines to invert, at least one MPI worker
rank will never receive a scanline index assignment it is waiting for, and the code will
\textit{never} successfully exit.  Since that MPI worker rank will perpetually wait for an
assignment that will never come, it will never be available to receive the termination message
from the MPI supervisor rank, and will hang the program execution.  Hence, this warning message and
exit condition are needed just to be safe.

\clearpage
\section{Code Functionality}

\subsection{Parameter File}\label{s:params}
The file \texttt{params.f90} contains constant parameters and global variable declarations
used throughout the VFISV code.  Any module requiring access to a parameter or global variable
must include the line \texttt{USE PARAMS} in the module header.  For example, the parameter
file contains declarations for the following quantities (among many others):\\

\begin{itemize}
  \renewcommand\labelitemi{\scriptsize$\blacksquare$}
  \item Fortran 90 variable kind types and numerical constants, e.g.:\\
        \texttt{INTEGER, PARAMETER :: DP = KIND(1.0D0)}
  \item Physical constants, e.g.:\\
        \texttt{REAL(DP), PARAMETER :: DPI = 3.141592653589793238462643\_DP}
  \item Runtime constants, e.g.:\\
        \texttt{INTEGER, PARAMETER :: NUM\_SCANS = 2048}
  \item Input/output configuration, e.g.:\\
        \texttt{INTEGER, PARAMETER :: NUM\_LEV1\_MEMAPS = 12}
  \item Quicklook and inversion engine parameters, e.g.:\\
        \texttt{INTEGER, PARAMETER :: MAX\_ITERS = 100}
  \item Line-formation \& spectral parameters, e.g.:\\
        \texttt{REAL(DP), PARAMETER :: LAB\_LAMBDA0 = 6302.4995\_DP}
  \item Global variables that are initialized at runtime, e.g.:\\
        \texttt{CHARACTER(LEN=256), DIMENSION(NUM\_SCANS) :: OBSFILES}
\end{itemize}

\subsection{Main Program}\label{s:main}
The main VFISV program is nothing more than an initialization, bookkeeping,
and collection routine which dispatches scanline inversion assignments to MPI worker ranks and
collects/assembles the results received back from the MPI worker ranks.  There are 
three main tasks coordinated by the main VFISV program: initialization, non-magnetic Stokes
\textit{I} profile calculation, and supervisor-worker inversion workflow, which are detailed in the following
subsections.

\subsubsection{Initialization}\label{ss:init}

The first step in the program is to perform sanity checks on command line input.\\

\textbf{SANITY CHECK \#1:}\\
The given Level-0.5 obs directory (\texttt{INDIR}) and Level-1 output directory (\texttt{WORKDIR})
are tested for existence.  If \texttt{INDIR} is not present on the command line or does not exist,
VFISV exits with an error.  If \texttt{WORKDIR} is not present on the command line, VFISV
exits with an error.  However, if it is present on the command line, but does not exist in the
filesystem, a system call will create it.  Note that the test for existence relies on the
\texttt{INQUIRE} function, used with the \texttt{ifort} extension argument \texttt{DIRECTORY}
to set the boolean flag \texttt{DIR\_EXISTS}:\\
\\
\texttt{INQUIRE( DIRECTORY=INDIR, EXIST=DIR\_EXISTS )}.\\
\\
The use of this extension makes the VFISV code system-dependent and non-portable.  
Once \texttt{INDIR} is validated, the directory is probed to determine the spatial
size of the dataset it contains.  The subroutine \texttt{DATA\_INIT()} is called, which
determines the number of files matching the \texttt{"k?v9s*"} regular expression and
sets the global variable \texttt{NUM\_FILES} equal to this number.  One of the Level-0.5
FITS files contained in this directory is then opened and used to determine the number of spatial
pixels along the scanline, \texttt{NUM\_PIX}.  The various arrays which are dependent on
the spatial dimensionality of the dataset are then dynamically allocated.  This allows the inversion
code to handle datasets of arbitrary spatial dimensionality, so that the code is not
limited to inversions of 2048 scanlines, each containing Stokes spectra for 2048 pixels.  In
the \solisvsm production pipeline, \texttt{INDIR} typically points to a directory containing
defringed spectra in an obs' tmpdir, for example:\\
\\
{\small \texttt{/path/to/vsm/tmp/k?v90YYMMDDtHHMMSS/k?v9sYYMMDDtHHMMSS\_oidXXXXXXXXXXXHHMM\_cleaned}}.\\
\\
However, when running in ``preview mode'', \texttt{INDIR} will point instead to the directory
containing the non-defringed spectra, for example:\\
\\
{\small \texttt{/path/to/vsm/tmp/k?v90YYMMDDtHHMMSS/oidXXXXXXXXXXXHHMM}},\\
\\
and some special manipulation is used to properly set the value of \texttt{OBSID} for the \texttt{DATA\_INIT()}
subroutine.\\

\textbf{SANITY CHECK \#2:}\\
If provided on the command line, \texttt{SCAN\_START} and \texttt{SCAN\_END} are checked to ensure
they are both positive and less than the maximum number of scanlines/pixels (typically 2048),
determined from \texttt{DATA\_INIT()}.  If not present, VFISV defaults to a full-disc inversion.\\

Once the command line input has been validated, runtime properties of the inversion
code can be initialized:\\

\textbf{CAMERA TYPE INITIALIZATION:}\\
The year of the observation is extracted from \texttt{INDIR}.  The value of the 
\texttt{CAMTYPE} variable is set to \texttt{ROCKWELL} for pre-2010 observations, or
to \texttt{SARNOFF} for observations from 2010 to present.  This variable allows
for easy discrimination in the code between the two camera eras, and is required for the
following:

\begin{itemize}
  \renewcommand\labelitemi{\scriptsize$\blacksquare$}
  \item Discrimination between on-disc and sky pixels, which uses a different
        intensity threshold for each camera era.  The on-disc intensity threshold for 
        the Rockwell camera (2003-2009) is set at 1500 counts, while for the Sarnoff
        camera (2010-present) the threshold is 5000 counts.
  \item The continuum window (range of wavelengths for which the continuum intensity
        is calculated) is different for the two cameras, due to the different
        illuminated portions of the camera CCDs.  The same is true for the individual 
        line windows (range of wavelengths which approximately centrally-bracket both
        the solar and telluric lines).
\end{itemize}

NOTE: any other camera-specific quantities can be set within the same routine that
performs the above initializations (\texttt{SET\_CAMTYPE\_PROPERTIES()}).\\

\textbf{DATA FILE INITIALIZATION:}\\
Once the input and output directories have been validated, the subroutine \texttt{FILE\_INIT()}
inspects the contents of the \texttt{INDIR} directory, and looks for standard \solisvsm 6302v
Level-0.5 FITS files, again using \texttt{INQUIRE} to test for existence.  For
bookkeeping purposes, \textit{all} FITS files in the input directory are listed
in the \texttt{OBSFILES(:)} array, but only those which are to be inverted (which have
an index between \texttt{SCAN\_START} and \texttt{SCAN\_END}) are listed
in the \texttt{INVFILES(:)} array.  This is for proper handling of cases where only
a subset of the observed scanlines are being inverted, as well as for handling
inversions of true area-scan observations.\\

\textbf{ME MODEL INITIALIZATION:}\\
Model parameters which are considered free parameters of the fit are marked as
such in the \texttt{FREE(:)} array, which is a boolean array with \texttt{.TRUE.} entries
for free parameters and \texttt{.FALSE.} entries for fixed parameters.  This allows the user
to invert for any combination of free and fixed parameters, and the algorithm will automatically
scale the sizes of its working spaces accordingly.  In the current implementation, only $\alpha$
and $v_{\mathrm{mac}}$ are considered fixed parameters, set to a respective constant value for
the entire run of the inversion (see Section \ref{ss:ql}).  In addition, if any regularization
of the free model parameters is required, this can be specified in the same manner, by setting the
corresponding entry of the \texttt{REGUL\_FLAG(:)} array to \texttt{.TRUE.}, and specifying the 
functional form of the regularization in the subroutine \texttt{REGULARIZE()}.

\subsubsection{Non-magnetic Stokes \textit{I} Profile Calculation}\label{ss:nonmag}
The non-magnetic Stokes \textit{I} profile used in the inversion is calculated as an average Stokes
\textit{I} profile over a local neighborhood around each pixel.  A 25 $\times$ 25 pixel$^{2}$ neighborhood
is used, centered on each pixel.  Only pixels with polarization degree, $p$, given by:

\begin{equation}
p = \frac{\mathrm{\mathbf{maxval}}\left(\sqrt{Q_{\lambda}^2 + U_{\lambda}^2 + V_{\lambda}^2}\right)}{I_{c}}
\end{equation}

less than $10^{-2}$ are allowed to contribute to the neighborhood average.  There can be situations where no
neighborhood pixels satisfy this criterion, for example in the deep umbra of larger sunspots.
In this situation, there would be no non-magnetic Stokes \textit{I} profile available for these pixels.
Simply increasing the size of the window so that (say) the largest anticipated magnetic structure
still has some neighboring quiet-Sun pixels is computationally inefficient and leads to prohibitively
long runtimes.  To alleviate this problem, a simple linear interpolation between neighboring profiles
is used to fill in these gaps.  Note that the non-magnetic Stokes vector is assumed to be unpolarized,
i.e., there is no non-magnetic Stokes \textit{Q}, \textit{U}, or \textit{V} signal.  The non-magnetic
profile is used in several support roles prior to the actual inversion itself, in which the non-magnetic
profile is synthesized from the same model parameters as the magnetic profile (but with no magnetic
field; see Section \ref{s:syn}).

\subsubsection{Supervisor-Worker Inversion Workflow}\label{ss:m-s}
\begin{figure}[!ht]
\centering
\includegraphics[width=\textwidth]{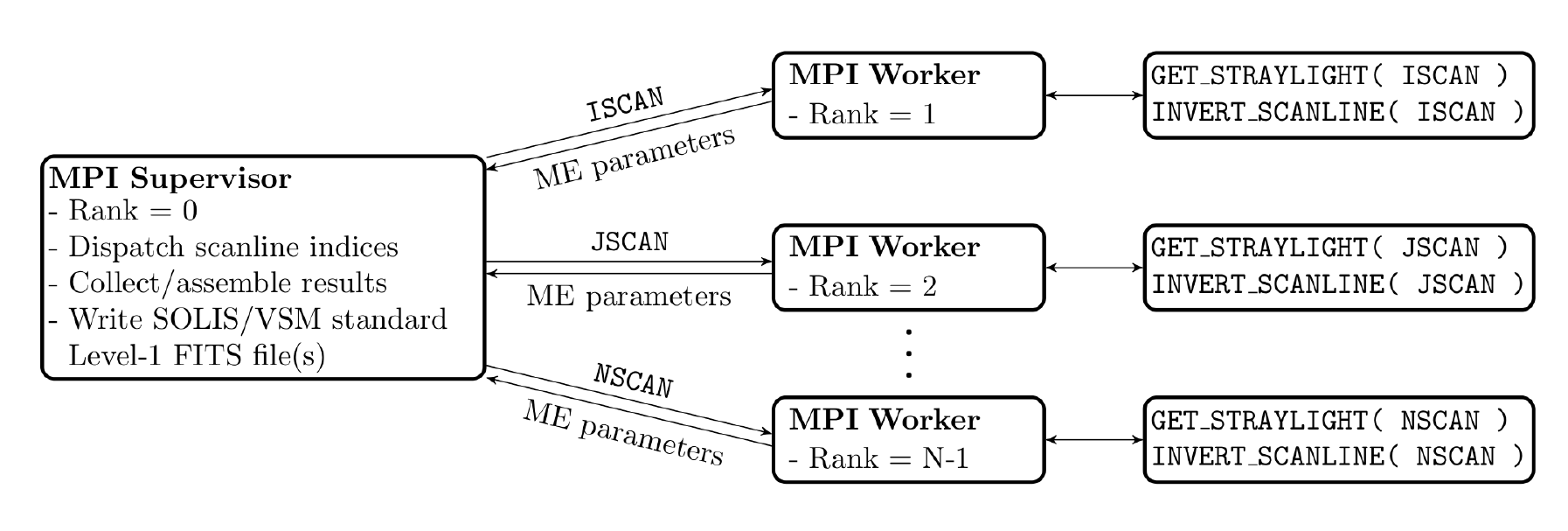}
\caption{Schematic flowchart describing division of labor amongst MPI ranks.}
\label{fig:mpi_flow}
\end{figure}

The MPI supervisor rank (\texttt{MPI\_RANK=0}) is responsible for coordinating the workflow amongst
the other MPI worker ranks, which are themselves responsible for performing the actual inversion of
the Stokes profiles in each \solisvsm Level-0.5 scanline.  To facilitate this coordination,
the MPI supervisor rank must be aware of which MPI worker rank is inverting which scanline, and
where the results should be placed in the resulting output array(s).  The MPI supervisor rank
does this by dispatching the scan index of a scanline to be inverted to each MPI worker rank,
and receiving the inverted parameters back from that worker rank, as shown schematically in
Figure \ref{fig:mpi_flow}.  When receiving results from an MPI worker rank (using the 
\texttt{MPI\_RECV()} subroutine), an array of supplementary information is also passed back
via the \texttt{MPI\_STAT(:)} argument.  \texttt{MPI\_STAT(MPI\_STATUS\_SIZE)} is an array
of integers containing the source rank, tag, and error code of the received message.
\texttt{MPI\_SOURCE}, \texttt{MPI\_TAG}, and \texttt{MPI\_ERROR} are the indices of the entries
for the source rank, tag, and error codes of the received message.  By setting these values, each
MPI worker rank can effectively communicate back to the MPI supervisor rank where the results should
be placed in the output array, as well as indicating that it is immediately available to invert another scanline.
On average, this speeds up the algorithm by about 15\% over a sequentially-ordered dispatch-and-return
scheme.  The following outline details how the MPI supervisor rank coordinates the workflow of the
MPI worker ranks until the entire dataset has been inverted:

\begin{itemize}
  \renewcommand\labelitemi{\scriptsize$\blacksquare$}
  \item The MPI supervisor rank dispatches the initial batch of scanline indices to worker ranks.
  \item Each MPI worker rank receives a scanline index, calculates the non-magnetic Stokes
        \textit{I} profiles for that scanline, and performs the ME inversion on the Stokes profiles
        in that scanline.  
  \item Each MPI worker rank dispatches the ME results back to the MPI supervisor rank, setting
        \texttt{MPI\_STAT(MPI\_TAG)} equal to the scanline index it initially received from the
        MPI supervisor rank.
  \item The MPI supervisor rank is always listening, and accepts results from any MPI worker rank,
        and uses the MPI tag located in \texttt{MPI\_STAT(MPI\_TAG)} to place the received results
        in the correct place in what will eventually be the output data product array written to an
        external FITS file.
  \item The MPI rank of the sender is identified by the value of \texttt{MPI\_STAT(MPI\_SOURCE)},
        so that the MPI supervisor rank knows which MPI worker rank it has just received results from, 
        and can dispatch another scanline index to that MPI rank.
  \item The MPI supervisor rank then dispatches the next available scanline index to that same
        MPI worker rank.  The cumulative number of scanline indices sent  is updated.
  \item This process repeats until results from all invertible scanlines have been received, and
        the total number of scanline indices dispatched (and number of resulting arrays received)
        by the MPI supervisor rank is equal to the number of scanlines to be inverted (\texttt{NUM\_SCANS}).
        The MPI supervisor rank then sends a special termination message to all MPI worker ranks to
        indicate there is no more work to be done, so that each MPI worker rank exits the loop in
        which it expects to receive a scanline index from the MPI supervisor rank.
  \item Note: all MPI subroutine calls are followed by a sanity check on the error flag, \texttt{IERR},
        set by the previously-called MPI subroutine.  This ensures correct operation and allows
        the inversion code to exit gracefully should anything go wrong with a call to an MPI subroutine.
\end{itemize}

The MPI supervisor rank rearranges the received output into an array that follows \solisvsm 6302v
Level-1 standards, then writes it to an external FITS file with a minimal header (which will later
be populated downstream in the Level-1 to Level-2 processing).  More detail on the output format is
given in Section \ref{ch:output}.  Runtime and inversion statistics are calculated and reported by
the MPI supervisor rank, then all MPI ranks synchronize and exit.  At this point, the VFISV inversion
is complete.

\subsection{Inversion Routines}\label{s:invert}
\begin{figure}
\centering
\includegraphics[width=0.65\textwidth]{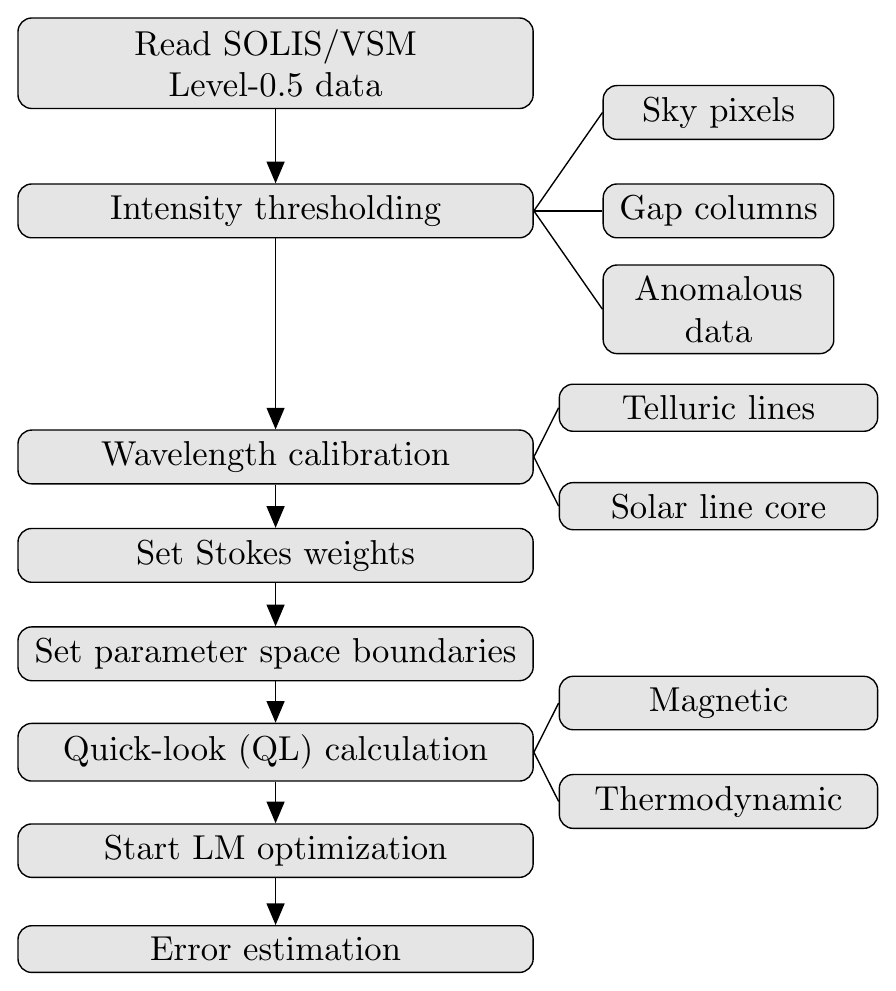}
\caption{Schematic flowchart showing the major steps of the inversion subroutine, \texttt{INVERT\_SCANLINE()}.}
\label{fig:inv_flow}
\end{figure}

This Section outlines the main subroutines used in the actual inversion subroutine,
\texttt{INVERT\_SCANLINE()}.  For reference, Figure \ref{fig:inv_flow} shows a schematic
flowchart of the major steps involved in the inversion of a \solisvsm 6302v Level-0.5 scanline.
Each major block in the flowchart is the subject of a subsection below.

\subsubsection{Reading the Data}\label{ss:data}
VFISV relies on the CFITSIO library (and the FORTRAN extensions therein) for reading and writing FITS files.
Once an MPI worker rank receives the scanline index for an inversion, the first step is to read that
scanline data file.  A free logical unit number is retrieved and associated with the file.
The file is opened, and the image dimensions are read from the FITS header (via the \texttt{NAXIS}
keywords).  A buffer of suitable size is allocated to hold the data, and the data is read
from the FITS file.  Upon output, the data in the buffer is converted to double precision.
The positions of the gap columns (due to the presence of the beamsplitter used to direct
solar light to the two \solisvsm cameras) are also read from the appropriate FITS header 
keywords (\texttt{GAPCOL1} and \texttt{GAPCOL2}).

\subsubsection{Intensity Thresholding}\label{ss:thresh}
While VFISV is meant to invert every single pixel in the field-of-view (FOV), some concessions
must be made about what qualifies as truly invertible data.  Before any inversion calculations
are made for a particular pixel, the code determines whether it should actually attempt to do
so, based on intensity considerations.  The following list outlines the situations in which no
inversion is attempted for a given pixel:

\begin{itemize}
  \renewcommand\labelitemi{\scriptsize$\blacksquare$}
  \item \textbf{Sky pixels:} These are low-intensity pixels outside the image-plane solar disc.
        Intensity thresholds for ignoring these pixels are different for Rockwell and Sarnoff
        camera eras, and are set in the \texttt{SET\_CAMTYPE\_PROPERTIES()} routine, as outlined
        in Section \ref{ss:init}.
  \item \textbf{Gap columns:} These are pixels (near the horizontal center of the FOV) containing
        non-invertible data resulting from the presence of the beamsplitter which directs solar
        light to camera A/B.  \texttt{GAPCOL1} and \texttt{GAPCOL2} header keywords are read from
        the Level-0.5 FITS file, and denote the starting and ending indices, respectively, of this
        gap in the data.  No thresholding based on intensity is performed in the gap, and the
        inversion code simply skips over pixels in the gap.
  \item \textbf{Anomalous data:} These pixels are likely the result of ``garbage'' at the
        extreme edges of the FOV; they may be straggler sky pixels, or the result of errors in
        the camera CCD readout, or something else.  If (1) the maximum total polarization signal
        is equal to zero, and/or (2) the continuum intensity of the neighborhood-averaged non-magnetic
        Stokes \textit{I} profile is zero, the inversion code simply skips over these pixels.
\end{itemize}

\subsubsection{Wavelength Calibration}\label{ss:wave}
Once the spectra in a given pixel have been validated as containing invertible data, the next
step is to calibrate the observed wavelength scale.  This is done by taking advantage of the 
two nearby terrestrial absorption lines of molecular oxygen.  They have a known wavelength
separation, so that we can establish the spectral dispersion by finding their positions
to subpixel accuracy.  Locating the telluric line-core region (to the nearest pixel) is 
done with a quasi-dynamic watershed algorithm, in which a suitably large window surrounding
each telluric line is searched in a ``geographic'' manner to find the minimum intensity.
Consider the line profile as a topological surface; the watershed algorithm adds ``droplets''
at each wavelength point in the telluric window.  Each droplet travels downhill (in the
negative gradient direction) until it encounters a local minimum, where it stops.  The
wavelength point which has accumulated the most droplets is marked as the core of the
telluric line.  This has distinct advantages over a simple search for the minimum intensity
of the line.  Since the telluric lines do not shift due to solar rotation, using
a large range of wavelengths when considering telluric line boundaries can ``bleed'' into
the solar line.  It is possible (and indeed happens) that there can be a wavelength point
in the solar line at a lower intensity than the telluric line-core, which would falsely
assign a solar line wavelength as the telluric line-core.  The watershed skirts this issue
by allowing \textit{some} misidentifications without influencing the final identification.
Some droplets may indeed fall into the solar line, but the majority of the droplets will
find their way into the telluric line.  This method is also insensitive to any
global shifts in the spectrum which might cause the telluric line(s) to wander outside
the (constant, hard-wired) telluric window.  Should the identified line-core position
fall outside the telluric window, the watershed algorithm recursively calls itself
with a smaller telluric window until the identified minimum falls within that window.
This method has proven to be extremely robust.\\

Once the telluric line-cores have been identified, the line-core regions are fit with a
second degree polynomial to refine the line-core positions to sub-pixel accuracy.  If
the two telluric line-core positions (in fractional pixels) are denoted $s_{1}$ and $s_{2}$
respectively, then the spectral dispersion, $\Delta\lambda_{\mathrm{disp}}$, is given by:

\begin{equation}
\Delta\lambda_{\mathrm{disp}} = \frac{\Delta\lambda_{\mathrm{O_{2}}}}{s_{2}-s_{1}},
\end{equation}

where $\Delta\lambda_{O_{2}}$ is the known telluric line separation.  This quantity (in \AA)
is given in the file \texttt{params.f90} as \texttt{DLAMO2 = 0.76220703D0}.  With a 
known spectral dispersion, and the position of one of the telluric lines known to be
6302.0005\AA, a calibrated wavelength offset scale can be obtained for each line.  This wavelength
offset scale is expressed as a shift from the line-core position of each line.  The line-core positions for the
solar lines are found by smoothing the wavelength derivative of Stokes \textit{I} with a Gaussian
kernel, then searching for the zero-crossing of the (smoothed) result.  Consistency is maintained
by allowing the 6302.5\AA\ line-core to vary a little around a 41-pixel offset from the core
of the 6301.5 line.  This helps prevent misidentifications due to (e.g.) highly split lines, or 
some anomalous data in the line profile.  Centering the wavelength scale on the solar line-core
means that positions on the line profile to the blue-side of the line-core have negative shifts, and
red-side line profile positions have positive shifts.  This transforms the reduced wavelength
array ($v$ in Equation (\ref{eq:v})) used in the Voigt- and Faraday-Voigt profile synthesis to:

\begin{equation}
v = \frac{\lambda-\lambda_{0}(v_{\mathrm{los}}/c) \pm \Delta\lambda_{Z}}{\Delta\lambda_{D}}.
\end{equation}

Since this is done only once in the establishment of the wavelength shift array, it is much
more efficient than using arrays of absolute wavelengths, and saves many repeated
calculations in the \texttt{SYNTHESIS()} subroutine.

\subsubsection{Parameter Space Boundaries}\label{ss:bounds}
The boundaries of the parameter space used in the inversion must be suitably large
enough to capture the range of physical atmospheric conditions present in the solar
photosphere, but restricted enough to allow the inversion to converge to a solution
in a reasonable amount of time.  The ME model parameters used in the inversion are
subject to the general box constraints shown below.  If any model parameter violates
its upper/lower box constraints at any point during the inversion, it is reset to the
corresponding upper/lower value.

\begin{align}
\left[\mathrm{min}(B),\mathrm{max}(B)\right] &= [0,3500]\ \mathrm{G}\\
\left[\mathrm{min}(\gamma),\mathrm{max}(\gamma)\right] &= [0,180]^{\circ}\ \ \ \ \ \star\\
\left[\mathrm{min}(\chi),\mathrm{max}(\chi)\right] &= [0,180]^{\circ}\\
\left[\mathrm{min}(v_{\mathrm{los}}),\mathrm{max}(v_{\mathrm{los}})\right] &= [-7 \times 10^{5},7 \times 10^{5}]\ \mathrm{cm/sec}\ \ \ \ \ \star\\
\left[\mathrm{min}(\Delta\lambda_{D}),\mathrm{max}(\Delta\lambda_{D})\right] &= [10,65]\ \mathrm{m\AA}\\
\left[\mathrm{min}(a_{\mathrm{dc}}),\mathrm{max}(a_{\mathrm{dc}})\right] &= [0,5]\\
\left[\mathrm{min}(\eta_{0}),\mathrm{max}(\eta_{0})\right] &= [1,100]\\
\left[\mathrm{min}(S_{0}),\mathrm{max}(S_{0})\right] &= [0,I_{c}]\ \ \ \ \ \star\\
\left[\mathrm{min}(\mu S_{1}),\mathrm{max}(\mu S_{1})\right] &= [0,I_{c}]\\
\left[\mathrm{min}(\alpha),\mathrm{max}(\alpha)\right] &= [0,1]\\
\left[\mathrm{min}(v_{\mathrm{mac}}),\mathrm{max}(v_{\mathrm{mac}})\right] &= [1 \times 10^{-5},5 \times 10^{5}]
\end{align}

Parameters marked by a $\star$ above are dealt with a little more rigorously, as follows:

\begin{itemize}
  \renewcommand\labelitemi{\scriptsize$\blacksquare$}
  \item Instead of using constant boundaries for the inclination angle, $\gamma$, which encompass
        both positive and negative polarities, the order of the Stokes \textit{V} lobes is used to
        constrain the parameter space; if the positive Stokes \textit{V} lobe is blueward of the
        negative Stokes \textit{V} lobe, the inclination is positive, and constrained to $[0,90]^{\circ}$.
        Conversely, if the negative lobe is blueward of the positive lobe, the inclination is
        thus constrained to $[90,180]^{\circ}$.  Quiet-sun pixels are liberally defined as those pixels
        whose quicklook magnetic filling factor, $\alpha$ (see Section \ref{ss:ql}), is less than 0.025,
        and for these pixels the entire inclination range is used ($\gamma \in [0,180]^{\circ}$).
  \item Since we have already accurately determined the line-center wavelength, we limit the range of
        allowable LOS velocities to values such that the line is only allowed to be Doppler shifted by
        a maximum of $\pm 1$ spectral pixel.
  \item As shown in Section \ref{s:me-model}, the source function parameters are constrained
        by $S_{0} + \mu S_{1} = I_{c}$.  To satisfy this relation in the inversion, only the
        source function gradient, $\mu S_{1}$, is considered a free parameter of the fit (since
        it affects the amplitudes of the Stokes profiles), and $S_{0}$ is then always calculated
        from the constraint relation.
\end{itemize}

\subsubsection{Quicklook (QL) Calculation(s)}\label{ss:ql}
The quicklook (QL) algorithm produces both a unique data product as well as an initialization
for the ME inversion.  It estimates the field strength, inclination, azimuthal angle, and magnetic
filling-factor directly from the Stokes profiles, integrated over suitable wavelength ranges.
The method is based on the technique in \citet{ronan:1987}, whereby Taylor expansions
of the relationship between intensity and circular/linear polarization profiles (for a normal Zeeman
triplet) are utilized to calculate the longitudinal/transverse components of the magnetic field.  Let
us define the integrated Stokes profiles as:

\begin{align}
\langle I \rangle &= \dint_{\lambda_{L}}^{\lambda_{U}}I_{\lambda}\mathrm{d}\lambda\\
\langle Q \rangle &= \dint_{-\infty}^{\lambda_{L}}Q_{\lambda}\mathrm{d}\lambda -
                     3.5\dint_{\lambda_{L}}^{\lambda_{U}}Q_{\lambda}\mathrm{d}\lambda +
                     \dint_{\lambda_{U}}^{-\infty}Q_{\lambda}\mathrm{d}\lambda\\
\langle U \rangle &= \dint_{-\infty}^{\lambda_{L}}U_{\lambda}\mathrm{d}\lambda -
                     3.5\dint_{\lambda_{L}}^{\lambda_{U}}U_{\lambda}\mathrm{d}\lambda +
                     \dint_{\lambda_{U}}^{-\infty}U_{\lambda}\mathrm{d}\lambda\\
\langle V \rangle &= \dint_{-\infty}^{\lambda_{0}}V_{\lambda}\mathrm{d}\lambda -
                     \dint_{\lambda_{0}}^{\infty}V_{\lambda}\mathrm{d}\lambda,
\end{align}

where $\lambda_{0}$ is the line-center wavelength.  The quantity
$\lambda_{UL} = \lambda_{0} \pm 1.5\Delta\lambda_{\mathrm{FWHM}}$, where $\Delta\lambda_{\mathrm{FWHM}}$
is the full width at half-maximum of the quiet-Sun Stokes \textit{I} profile.  Although some of the
upper/lower limits above are $\pm\infty$, in practice any limit which reaches the (unpolarized) continuum
will give the same results.\\

With the above definitions, the LOS and transverse (TRN) components of the magnetic field are given by:

\begin{align}
B_{\mathrm{LOS}} &= C_{\mathrm{LOS}}\left[\frac{\langle V \rangle}{\langle I \rangle}\right]\label{eq:blos}\\
B_{\mathrm{TRN}} &= C_{\mathrm{TRN}}\left[\left(\frac{\langle Q \rangle}{\langle I \rangle}\right)^{2} + 
                       \left(\frac{\langle U \rangle}{\langle I \rangle}\right)^{2}\right]^{\frac{1}{4}},\label{eq:btrn}
\end{align}

from which the QL inclination angle, $\gamma$, can be derived as:

\begin{equation}
\gamma = \mathrm{tan}^{-1}\left(\frac{B_\mathrm{TRN}}{B_{\mathrm{LOS}}}\right).
\end{equation}

The QL transverse azimuthal angle, $\chi$, can be similarly calculated from:

\begin{equation}
\chi = \frac{\pi}{2} - \frac{1}{2}\mathrm{tan}^{-1}\left[\frac{-{\langle U \rangle}}{{\langle Q \rangle}}\right]
\end{equation}

The QL magnetic filling-factor, $\alpha$, is calculated as in \citet{bommier:2009}:

\begin{equation}
\alpha = \frac{ \left[V_{\lambda}^{2}\right]_{\mathrm{max}}\mathrm{tan}^{2}\gamma}
              {\left[Q_{\lambda}^{2} + U_{\lambda}^{2}\right]_{\mathrm{max}}\left(I_{c}-I_{0}\right)},
\end{equation}

where $I_{0}$ is the line-core intensity.  We choose not to consider the filling-factor as a free
parameter of the fit, but instead hold it fixed at the QL value for the duration of the inversion.
Given the well-known degeneracy between magnetic field strength and filling-factor when trying to 
invert for them simultaneously, as well as the fact that the filling-factor is really a kind of zeroth-order
attempt to account for sub-pixel structure, we do not consider this too restrictive
of an assumption, especially considering the quality of the results produced by the QL filling-factor
calculation.\\

The proportionality constants $C_{\mathrm{LOS}}$ and $C_{\mathrm{TRN}}$ have been determined
\textit{a priori} by linear regression between the LOS and transverse components as determined
from \sdohmi observations on a test dataset and the corresponding quantities multiplying the 
proportionality constants in Equations (\ref{eq:blos}) and (\ref{eq:btrn}).  The \sdohmi magnetograms
were rebinned and smoothed using a Gaussian kernel to match the size and spatial sampling of \solisvsm.
This was accomplished by maximizing the cross-correlation between the \solisvsm and \sdohmi magnetograms
as a function of (e.g.) smoothing kernel size.  As implemented in the VFISV code, the proportionality
constants were found to have the values:

\begin{align}
C_{\mathrm{LOS}} &= 1.6833 \times 10^{4}\\
C_{\mathrm{TRN}} &= 8.6960 \times 10^{3}
\end{align}

It should be noted that these coefficients may need to be recalculated, particularly if any
changes are made to the Lev0 code, if any hardware is swapped, or hardware settings changed,
since these coefficients are derived from spectra with a given (e.g.) exposure time, frame
rate, etc.\\

Thermodynamic model parameters are initialized in a less-rigorous way, simply to obtain decent
starting values for the inversion; an approximate temperature is determined by referencing the 
continuum level to a nearby continuum wavelength in the atlas spectrum of \citet{labs:1968},
and inverting the Planck function to solve for temperature.  This reference intensity is
$3.11 \times 10^{14}\ \mathrm{ergs\ cm^{-2}\ s^{-1}\ sr^{-1}\ cm^{-1}}$ at $\lambda_{\mathrm{ref}} = 6306.0$\AA.
From the temperature, we can then derive estimates of the Doppler width ($\Delta\lambda_{D}$),
damping parameter ($a_{\mathrm{dc}}$), and line-to-continuum opacity ratio ($\eta_{0}$).  The source
function continuum, $S_{0}$, is initialized to zero, and its gradient, $\mu S_{1}$, is set to $I_{c}$,
in accordance with the constraint on the sum of these two parameters.  LOS velocity is also initialized
to zero.

\subsubsection{Stokes Weights}\label{ss:weights}
The weights $w_{ij}$ entering into Equation (\ref{eqn:chi2}) are used to adjust the relative
importance of deviations between the synthesized and observed Stokes profiles.  Here, the
$j$ index is dropped from the weights; they are taken as constant over the spectral line,
but distinct for each of the four Stokes profiles.  After \textit{exhaustive} testing of various
different weighting schemes, the following weights have been adopted, since they produce the best
results in terms of synthetic fits to the observed line profiles while minimizing spurious
pixel-to-pixel variations:

\begin{align}
w_{I} &= \frac{1}{I_c}\\
w_{QUV} &= \left[\mathrm{\mathbf{maxval}}\left(\sqrt{Q_{\lambda}^{2}+U_{\lambda}^{2}+V_{\lambda}^{2}}\right)\right]^{-1}.
\end{align}

This has the effect of roughly equalizing the importance of deviations in Stokes \textit{I},
\textit{Q}, \textit{U}, and \textit{V} which contribute to the total $\chi^{2}$ merit function
value, provided there is adequate signal.  For example, consider the situation of a sunspot umbral
pixel at disc-center.  In this case, Stokes \textit{V} will have a strong signal due to the strong 
LOS field component, while Stokes \textit{Q} and \textit{U} will be dominated by noise.  Therefore,
$w_{QUV}$ will be close to the maximum absolute value of the Stokes \textit{V} profile.  Along with
the $w_{I}$ weight, this will ensure that deviations in Stokes \textit{I} and \textit{V} contribute
roughly equally to the $\chi^{2}$ value.  Conversely, since the Stokes \textit{V} amplitude is much
greater than that of Stokes \textit{Q} and \textit{U} (which are noise-dominated), application of
these weights to Stokes \textit{Q} and \textit{U} will drastically reduce the importance of their
contributions to the $\chi^{2}$ value.  Similar arguments can be made for any other combination of
strong/weak Stokes \textit{Q}, \textit{U}, and \textit{V} profiles.\\

In addition, to further decrease the importance of Stokes \textit{Q}, \textit{U}, and \textit{V}
profiles in the quiet-sun (which are noise-dominated), the Stokes \textit{Q}, \textit{U}, and \textit{V} weights are all
multiplied by a simple function of the QL filling-factor, $\alpha$.  This has the effect of retaining the
original weights in strong field pixels while further decreasing the importance of Stokes \textit{Q},
\textit{U}, and \textit{V} in quiet-Sun pixels, where these signals are dominated by noise.  It is
well known that the presence of noise in these Stokes profiles can lead an inversion to falsely
interpret some of that noise as real signal, which can in turn lead to artificially-high field strengths
and poor profile fits.  The modification of the Stokes weights by a function of the filling factor
is an effective way to decrease the likelihood that noise-dominated Stokes profiles will contribute
to the fit in a meaningful way.  The functional form of this modification is:

\begin{equation}
f(\alpha) = \mathrm{\mathbf{min}}\left(\alpha+0.05,1\right), 
\end{equation}

such that $w_{QUV} \leftarrow f(\alpha)w_{QUV}$.
Note that $f(\alpha)$ has a floor of 0.025 and a ceiling of 1.  While in theory the floor should be 0
(to completely de-weight Stokes \textit{Q}, \textit{U}, and/or \textit{V} profiles which are solely
noise), I have noticed that this case tends to produce sharp boundaries in the inverted parameters at
the edge of magnetically active regions.  Having a small but non-zero floor smoothes out these
discontinuities satisfactorily.

\subsection{Stokes Profile Synthesis Module}\label{s:syn}
The synthesis code for calculating Stokes profiles takes the vector ME model parameters
and returns the Unno-Rachkovsky solutions to the PRTE.  The elements of the propagation
matrix, $\mathbf{K}_{\lambda}$, are evaluated and are used to calculate the Unno-Rachkovsky
solutions, $\mathbf{I}^{\mathrm{M}}_{\lambda}$, for both the 6301.5\AA\ and 6302.5\AA\ lines.
These synthesized lines are superimposed to create the full synthetic line profile in the 
appropriate wavelength range.  This full line profile is mixed with the non-magnetic
component, according to the value of $\alpha$.  We correct for scattered (or stray) light
in the observations at the 5\% level by adding the appropriate amount of scattered light
to the synthesized profiles (as opposed to removing the scattered light from the observed
spectra themselves before the inversion).  For a scattered light fraction, $s < 1$, the 
standard correction has the form:

\begin{equation}
I_{\mathrm{corr}} = \frac{I_{\mathrm{obs}} - s\langle I_{\mathrm{obs}} \rangle}{1-s}, 
\end{equation}

where $\langle I_{obs} \rangle$ denotes the average observed intensity in the line profile.  We
invert this expression to obtain a synthesized Stokes \textit{I} profile that has had the proper
amount of scattered light \textit{added} to it:

\begin{equation}
I_{\mathrm{syn}} = (1-s)I_{\mathrm{syn}} + s\langle I_{\mathrm{syn}} \rangle.
\end{equation}

\begin{figure}[!ht]
\centering
\includegraphics[width=\textwidth]{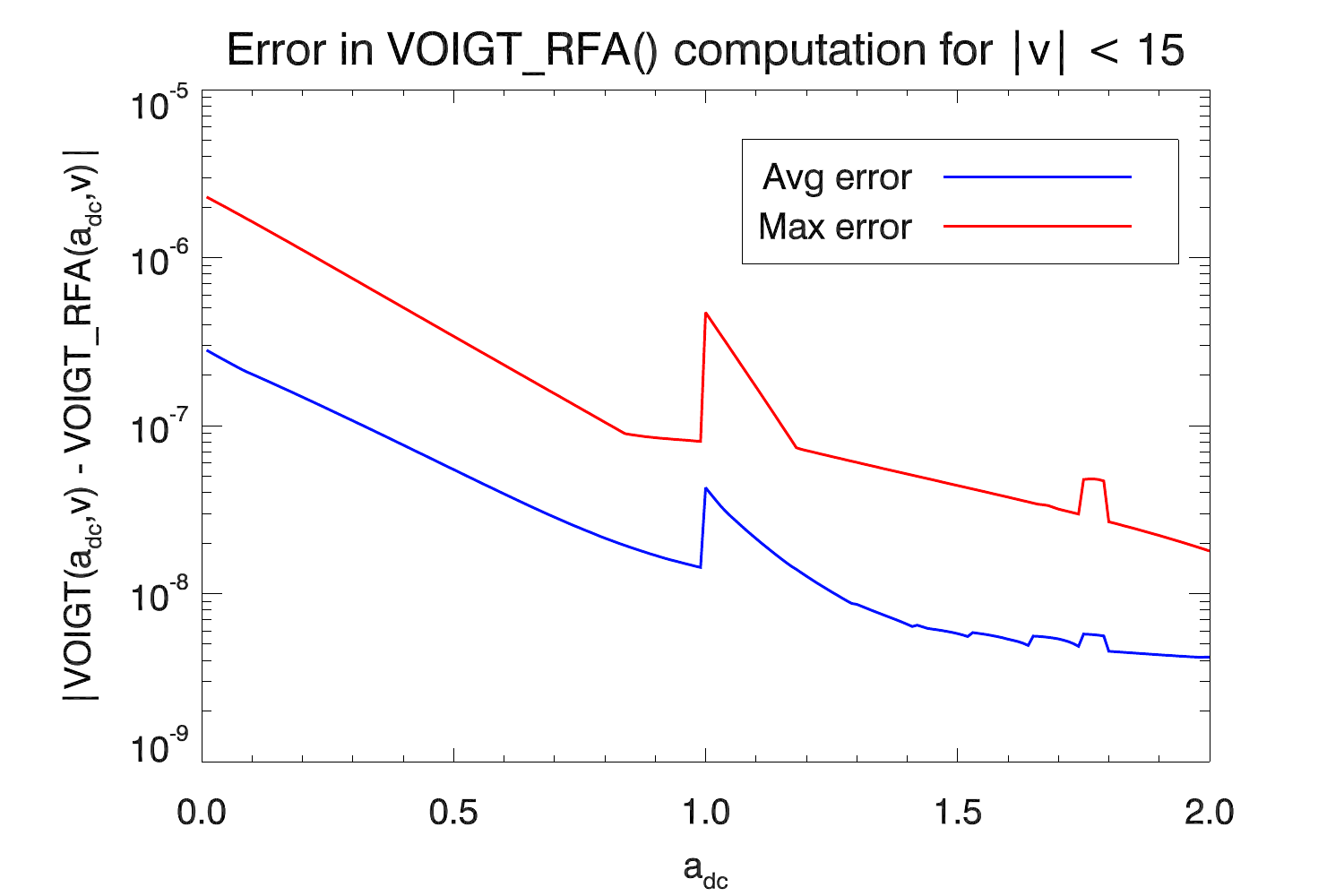}
\caption{Absolute errors between \texttt{VOIGT\_RFA()} and the IDL implementation.}
\label{fig:voigt}
\end{figure}

The 6302.5\AA\ line is a normal Zeeman triplet, with a single $\sigma_{r}$, $\sigma_{b}$, and
$\pi$-component; the 6301.5\AA\ line is an anomalous Zeeman triplet with multiple Zeeman subcomponents.
Each component depends non-linearly on the absorption ($\phi_{\lambda}$) and anomalous dispersion
($\psi_{\lambda}$) profiles, calculated from the Voigt and Faraday-Voigt lineshapes, respectively.
The \texttt{VOIGT\_RFA()} algorithm uses the rational function approximation (RFA) of \citet{hui:1978}.
Let

\begin{equation}
R(z) = \frac{A(z)}{B(z)}, 
\end{equation}

where $A(z)$ and $B(z)$ are complex polynomials of the complex number $z = a_{\mathrm{dc}} - v\mathrm{i}$.
Then $H(a_{\mathrm{dc}},v) = \mathrm{Real}[R(z)]$ and $F(a_{\mathrm{dc}},v) = \mathrm{Sign}(v)\mathrm{Imag}[R(z)]$.
This algorithm is very fast and accurate, and has been tested for accuracy against the built-in IDL 
function VOIGT().  Over the applicable domain of $a_{\mathrm{dc}}$ relevant for photospheric spectral line
inversion ($a_{\mathrm{dc}} \sim 0.5$), the two methods differ by at most an average of a few times
$10^{-7}$ in absolute value, as seen in Figure \ref{fig:voigt}.  The maximum error is about an order of 
magnitude greater than the average error, but these larger errors occur far out in the continuum, where
the computation has less effect on the line-shape.  The complex polynomial coefficients for $A(z)$ and $B(z)$
used in the RFA are defined in \texttt{params.f90} in the \texttt{ACOEFF} and \texttt{BCOEFF} variables.\\

The synthesis module also calculates the derivatives of the synthesized Stokes profiles with
respect to the model parameters, $\partial\mathbf{I}_{\lambda}^{\mathrm{M}}/\partial p_{i}$.  Under the
ME assumptions, these derivatives can be evaluated analytically.  To increase the speed of the
synthesis, and to avoid unnecessary calculations, these derivatives are evaluated only if the
$\chi^{2}$ merit function value has improved, except when a reset is performed (see Section \ref{ss:lm}),
in which case the derivatives \textit{must} be updated immediately.

\subsubsection{Levenberg-Marquardt Considerations}\label{ss:lm}
The QL model parameters are used as a starting point for synthesizing Stokes profiles; these 
synthesized profiles are compared to the observed profiles, and the $\chi^{2}$ goodness-of-fit
metric is evaluated.  The QL model parameters are then iteratively improved upon
by a modified Levenberg-Marquardt (LM) algorithm, by generating model parameter perturbations,
synthesizing new Stokes profiles, and judging whether the new profiles are an improvement over
the old profiles.  This proceeds until one or more termination criteria are met, at which point
the model vector corresponds to the model parameters of the ME atmosphere in which the observed
Stokes profiles were formed.  This iterative LM inversion is skipped entirely when running in
``preview mode''.\\

In order to make improvements to the fit between the observed and synthesized profiles, we need to
be able to generate perturbations in the model parameters in a self-consistent way.  This is done
by monitoring the curvature of the $\chi^{2}$ hypersurface, through the gradient vector and 
Hessian matrix.  Recalling Equation (\ref{eqn:chi2}), we can calculate the gradient vector and Hessian
matrix in a straightforward way.  The gradient with respect to the model parameters is therefore
given by

\begin{equation}
\left[\nabla\chi^{2}(\mathbf{p})\right]_{i} = \deriv{\chi^{2}(\mathbf{p})}{\mathrm{p}_{i}} =
-\frac{2}{\nu} \dsum_{k=IQUV}w_{k}^{2}\dsum_{N_{\lambda}}
\left[\mathbf{I}_{\lambda}^{\mathrm{O}_{k}}-\mathbf{I}_{\lambda}^{\mathrm{M}_{k}}\right]
\deriv{\mathbf{I}_{\lambda}^{\text{M}_{k}}}{\mathrm{p}_{i}},
\end{equation}

and the Hessian matrix of partial derivatives is given by

\begin{equation}
\left[\mathbf{H}(\mathbf{p})\right]_{ij} = \sderiv{\chi^{2}(\mathbf{p})}{\mathrm{p}_{i}}{\mathrm{p}_{j}} =
\frac{2}{\nu}\dsum_{k=IQUV}w_{k}^{2}\dsum_{N_{\lambda}}\left[\deriv{\mathbf{I}_{\lambda}^{\mathrm{M}_{k}}}{\mathrm{p}_{i}}
\deriv{\mathbf{I}_{\lambda}^{\mathrm{M}_{k}}}{\mathrm{p}_{j}}-\left[\mathbf{I}_{\lambda}^{\mathrm{O}_{k}}-
\mathbf{I}_{\lambda}^{\mathrm{M}_{k}}\right]\sderiv{\mathbf{I}_{\lambda}^{\mathrm{M}_{k}}}{\mathrm{p}_{i}}{\mathrm{p}_{j}}\right].
\end{equation}

It is conventional to disregard the second-derivative term in the above equation, since it may have a
destabilizing effect in the early stages of the optimization, when the deviations between the model
and observations are expected to be large.  At later stages (i.e. after improvements to the fit), the
contribution from this term will tend to cancel out when summed over wavelength, due to the presence of
the $\mathbf{I}_{\lambda}^{\mathrm{O}_{k}}-\mathbf{I}_{\lambda}^{\mathrm{M}_{k}}$ term.  Therefore, the
true Hessian matrix is approximated by the product of the corresponding first derivatives:

\begin{equation}
\left[\mathbf{H}(\mathbf{p})\right]_{ij} = \sderiv{\chi^{2}(\mathbf{p})}{\mathrm{p}_{i}}{\mathrm{p}_{j}} \approx
\frac{2}{\nu}\dsum_{k=IQUV}w_{k}^{2}\dsum_{N_{\lambda}}\left[\deriv{\mathbf{I}_{\lambda}^{\mathrm{M}_{k}}}{\mathrm{p}_{i}}
\deriv{\mathbf{I}_{\lambda}^{\mathrm{M}_{k}}}{\mathrm{p}_{j}}\right]
\end{equation}

Given the above equations describing the curvature of the $\chi^{2}$ hypersurface, we can relate them to
give a set of linear equations:

\begin{equation}\label{eqn:pert}
\mathbf{H}(\mathbf{p}) \cdot \Delta\mathbf{p} = \nabla\chi^{2}(\mathbf{p}),
\end{equation}

which may be solved for the model perturbations, $\Delta\mathbf{p}$.  Contributions from Stokes \textit{I} to
$\chi^{2}(\mathbf{p})$, $\nabla\chi^{2}(\mathbf{p})$, and $\mathbf{H}(\mathbf{p})$ for wavelengths within the
telluric window are neglected, to prevent these deviations from contributing to the solution procedure.  For 
stability and robustness, these linear equations are solved using the Singular Value Decomposition, as
implemented in the \texttt{DGESVD()} routine from the LAPACK library.\\

The LM algorithm uses an adaptive ``step-size'' governor parameter, $\lambda$, to allow the algorithm
to dynamically and smoothly switch itself between the quadratic approximation and gradient-descent paradigms,
when solving from the model parameter perturbations in each iteration; the diagonal elements of the Hessian
matrix are multiplied by the factor $1+\lambda$, so that

\begin{equation}
\left[\mathbf{H}(\mathbf{p})\right]_{ii} = \left(1 + \lambda\right)\left[\mathbf{H}(\mathbf{p})\right]_{ii}.
\end{equation}

\begin{figure}
\centering
\includegraphics[width=\textwidth]{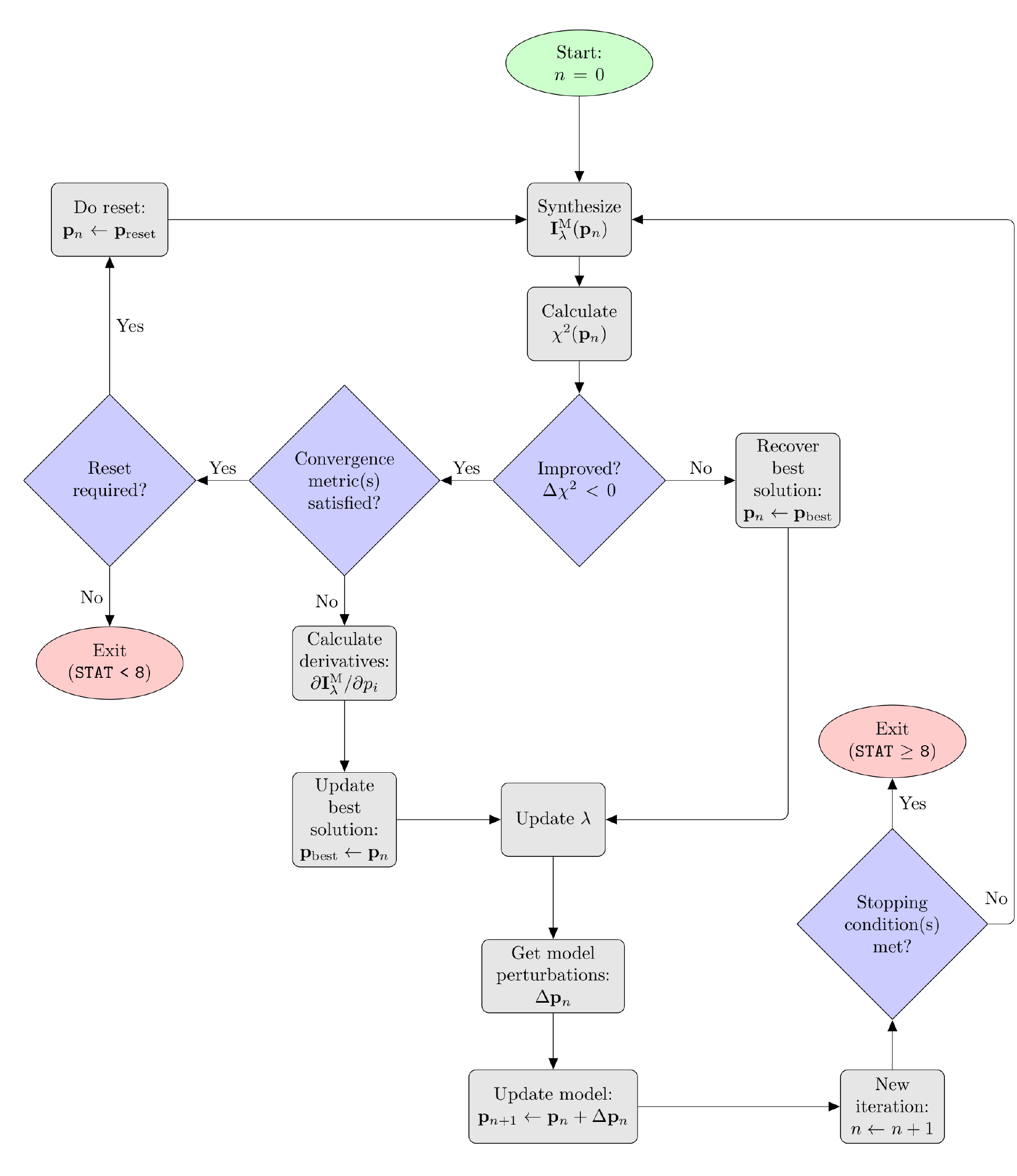}
\caption{Schematic flowchart detailing the operation of the LM optimization.}
\label{fig:lm_flow}
\end{figure}

For $\lambda \ll 1$, this modification has little effect, so that the solution to the above set of linear
equations is roughly the quadratic approximation, in which the neighborhood of the current vector $\mathbf{p}$
is approximated by a surface with quadratic curvature, and the model parameter perturbations calculated
from Equation (\ref{eqn:pert}) allow $\mathbf{p}$ to jump directly to the minimum of that surface. Conversely,
for $\lambda \gg 1$, the Hessian matrix becomes diagonally-dominant, and this has the effect of approximating
a gradient-descent algorithm when calculating the model parameter perturbations.  The value of the $\lambda$
parameter for the next iteration is adjusted according to the (non-)success of the current iteration:

\begin{equation}
\lambda_{n+1} = 
\begin{dcases}
\lambda_{n}/\epsilon_{-},& \text{if } \chi^{2}_{n} \leq \chi^{2}_{\mathrm{best}}\\
\epsilon_{+}\lambda_{n},& \text{if } \chi^{2}_{n} > \chi^{2}_{\mathrm{best}}
\end{dcases}
\end{equation}

where $\epsilon_{-} = 5$ and $\epsilon_{+} = 3$.  Historically, many algorithms set $\lambda_{n}$ to be increased
and decreased by the same factor.  The asymmetric adjustment used here is commonly known as ``delayed gratification'' and
has been shown to have better performance characteristics than the symmetric update; many different $\lambda$-update
strategies were tested during development, and this method did provide the best performance in terms of speed
of convergence and solution quality.  So, for a successful iteration in which an improvement in the $\chi^{2}$
goodness-of-fit metric is found, the model vector lies closer to the minimum of the $\chi^{2}$ hypersurface (i.e. closer
to the region where the quadratic approximation holds).  Therefore, we decrease the $\lambda$ factor, making the algorithm
behave more like it is calculating an exact solution for the minimum.  To prevent $\lambda$ from getting too small and
wasting computational effort, we set a minimum value, $\lambda_{\mathrm{min}} = 10^{-4}$.  For a non-successful iteration in
which no improvement is found, the $\lambda$ factor is increased and the algorithm attempts to traverse the parameter
space by taking ever-smaller steps in the negative gradient direction to improve the fit.  This procedure is iterated
until one or more convergence metrics are satisfied (more on this in Section \ref{ss:conv}).  A flowchart of the LM optimization
procedure is shown in Figure \ref{fig:lm_flow}.\\

Two ``insurance policy'' strategies are employed within the LM inversion algorithm, to ensure that
it does indeed converge to a good solution in the face of an unforseen anomalous runtime condition:
a recovery strategy and a reset strategy.  The details of these strategies are provided below.

\begin{itemize}
  \renewcommand\labelitemi{\scriptsize$\blacksquare$}
  \item \textbf{RECOVERY STRATEGY:} Should something go wrong at any point in the inversion (e.g. a
        \texttt{NaN} or \texttt{Inf} crops up), for whatever reason, a fallback mechanism is in place to re-establish a
        good point/direction in parameter space.  The downhill simplex method of \citet{nelder:1965},
        commonly known as the ``amoeba'' algorithm, is used.  The last known good solution (before the
        \texttt{NaN}/\texttt{Inf} showed up, e.g.) is used to randomly spawn a non-degenerate simplex, which
        then attempts to move in the negative gradient direction on the $\chi^{2}$ parameter space by ``reflecting''
        the highest vertex of the simplex through the opposite face of the simplex to a lower point.  Expansions and
        contractions of the vertex can also be performed, to take larger or smaller steps, respectively.  This continues
        until all vertices of the simplex have converged on each other to within some specified tolerance
        (defined in \texttt{params.f90} as \texttt{AMOEBA\_FTOL}).  The best vertex then overwrites the
        last known good solution, and the LM algorithm proceeds as normal.  To avoid too much time being spent
        in this recovery phase, the convergence tolerance for the amoeba algorithm is much larger than the normal $\chi^{2}$ convergence
        tolerance (see next Section), and a relatively small maximum number of amoeba iterations is used.
        The \solisvsm remix of the VFISV code is very stable and robust, so in practice this recovery strategy
        is \textit{very} rarely employed.  But, it never hurts to be prepared.
  \item \textbf{RESET STRATEGY:} Occasionally, some pixels do not find an adequate solution (i.e. $\chi^{2}$ does not
        improve ``enough'' over the initial estimate).  When this occurs, the model vector is re-initialized
        using the best-fit model from the previous pixel (along the scanline), various bookkeeping variables
        are re-initialized, and the inversion starts over with the new model vector.  In the case where multiple
        resets are required, the re-initialized model vector is perturbed randomly (the scale of the perturbations
        becomes proportionately larger for repeated resets).  This strategy ensures that any spurious pixel-to-pixel
        variation in the magnetograms (due to non-convergence of the solution trajectory) is minimized.  As an insurance
        policy to guard against a potential infinite loop, the maximum number of allowed resets is set in
        \texttt{params.f90} as \texttt{MAX\_RESETS=5}.
\end{itemize}

\subsubsection{Convergence Strategies/Criteria}\label{ss:conv}
The following conditions are used to signal successful convergence of the inversion algorithm, and 
the value of \texttt{CONVERGENCE\_FLAG} is set to indicate which convergence criterion was met.  To signal
successful convergence, we require any combination of the following conditions to be met for two (2) consecutive
successful iterations:

\begin{itemize}
  \renewcommand\labelitemi{\scriptsize$\blacksquare$}
  \item \textbf{CONVERGENCE CRITERIA 1:} An improvement produces a change in
        $\chi^{2}$ less than a specified convergence tolerance (\texttt{FCONV\_TOL}), defined in \texttt{params.f90}.
        In this case, \texttt{CONVERGENCE\_FLAG = 1} if no reset was triggered.  If a reset was triggered during
        the inversion, \texttt{CONVERGENCE\_FLAG = 5}.
  \item \textbf{CONVERGENCE CRITERIA 2:} An improvement produces a change in each model
        parameter (relative to the current best model) that is less than a specified convergence tolerance
        (\texttt{XCONV\_TOL}), defined in \texttt{params.f90}.  In this case, \texttt{CONVERGENCE\_FLAG = 2} if 
        no reset was triggered.  If a reset was triggered during the inversion, \texttt{CONVERGENCE\_FLAG = 6}.
  \item \textbf{CONVERGENCE CRITERIA 3:} The LM $\lambda$ parameter exceeds a maximum value (\texttt{LAMBDA\_MAX}), defined
        in \texttt{params.f90}.  This large $\lambda$ value indicates that the algorithm has
        spent many iterations in the ``gradient descent'' paradigm of the solution procedure,
        and therefore is likely in a low-curvature region of the parameter space, i.e. near a minimum in
        the parameter space.  The current implementation sets \texttt{LAMBDA\_MAX=1E+4\_DP}.  In this
        case, \texttt{CONVERGENCE\_FLAG = 3}. If a reset was triggered during the inversion,
        \texttt{CONVERGENCE\_FLAG = 7}.
\end{itemize}

In addition to the convergence criteria above, we must employ some stopping criteria to terminate the inversion
if the solution fails to converge in a timely fashion.  The two stopping criteria employed are as follows:

\begin{itemize}
  \renewcommand\labelitemi{\scriptsize$\blacksquare$}
  \item \textbf{STOPPING CRITERIA 1:} If the inversion has not signalled convergence after \texttt{MAX\_ITERS} iterations
        (defined in \texttt{params.f90}, default 200), the inversion stops iterating and reports the best solution it
        found.  In this case, we cannot assume the best-found model parameters are indeed the best-fit model
        parameters, and so any uncertainty calculation may not be meaningful.  If no reset was triggered during
        the inversion, \texttt{CONVERGENCE\_FLAG = 4}.  If a reset was triggered, \texttt{CONVERGENCE\_FLAG} = 8.
  \item \textbf{STOPPING CRITERIA 2:} If too many resets are triggered during the inversion, this is likely 
        a pixel with Stokes data of anomalous or strange morphology, and it is likely the inversion may not
        ever find a good solution.  The maximum number of allowable resets for a single pixel is specified by
        the \texttt{MAX\_RESETS} parameter (defined in \texttt{params.f90}, default 5).  If a pixel inversion fails
        due to too many resets, \texttt{CONVERGENCE\_FLAG = 9}.
\end{itemize}

\subsubsection{Error Estimation}\label{ss:errors}
Once the LM algorithm has converged, we can obtain estimates of the standard errors/uncertainties in the 
fitted model parameters.  To do this, the inverse of the Hessian matrix, known as the covariance matrix,
must be calculated.  This matrix describes the covariances between pairs of the fitted model parameters,
$\mathbf{p}_{\mathrm{best}}$.  The standard variances in the model parameters are proportional to the
diagonal elements of the covariance matrix, and are given by:

\begin{equation}
\sigma_{i}^{2} = \frac{\chi^{2}(\mathbf{p}_{\mathrm{best}})}{N_{\lambda}}\left[\mathbf{H}^{-1}(\mathbf{p}_{\mathrm{best}})\right]_{ii}.
\end{equation}

However, this standard definition relies on the assumption that, on exit, the gradient of $\chi^{2}$ is equal to
zero, and any uncertainties in the fit are due to noise in the observed spectra.  The LM algorithm typically does
not reach a zero-gradient solution (in a mathematical sense), so that these error estimates systematically
underestimate the true errors.  This is because the uncertainty in the fit may not be due solely to noise, but 
rather to the distance between the true minimum and the returned $\mathbf{p}_{\mathrm{best}}$ which has a 
value of $\chi^{2}({\mathbf{p}_{\mathrm{best}}})$.  \citet{sanchezalmeida:1997} instead developed an error estimate
which does not depend on the above assumptions, so is safe to evaluate the uncertainties in the model parameters
on exit.  It provides a range of free parameters whose synthetic Stokes profiles deviate from the observed
Stokes profiles by an amount equal to $\chi^{2}({\mathbf{p}_{\mathrm{best}}})$, thus providing a statistically
significant estimate of the uncertainty.  This new estimate turns out to be

\begin{equation}
\sigma_{i}^{2} = \frac{\chi^{2}(\mathbf{p}_{\mathrm{best}})}{N_{\mathrm{free}}}\left[\mathbf{H}^{-1}(\mathbf{p}_{\mathrm{best}})\right]_{ii},
\end{equation}

where $N_{\mathrm{free}}$ is the number of free parameters of the fit.  Note that, since $N_{\mathrm{free}} \ll N_{\lambda}$, we
are now no longer underestimating the errors.  These standard errors are only accurate
provided the final best model parameters are indeed close to the minimum of the $\chi^{2}$ hypersurface.
For cases where convergence to the minimum of the parameter space cannot be assumed (i.e. when the maximum
number of iterations were performed without signalling any convergence criteria), these errors are not
likely to be very meaningful.

\clearpage
\section{Output Format}\label{ch:output}

VFISV writes separate standard \solisvsm 6302v Level-1 data product FITS files containing both the
QL and ME output of the code.  However, when running in ``preview mode'', only the Level-1 QL is
is actually written to a file.  Regardless of the values of \texttt{SCAN\_START}, \texttt{SCAN\_END},
\texttt{PIX\_START}, and/or \texttt{PIX\_END} specified on the command line, these FITS files are
sized to 2048 $\times$ 2048 spatial pixels.  The FITS files are staged in temporary directories, to
be modified by later pipeline routines before being deployed to the keep.  The Level-1 ME(QL) file
contains 12(6) planes, and for completeness these planes are listed here in the order in which they
appear in the output file.  Each item is prefaced by an indication of which Level-1 data product it
appears in (ME and/or QL).

\begin{itemize}
  \renewcommand\labelitemi{\scriptsize$\blacksquare$}
  \item [\scriptsize$\blacksquare$] \textbf{[ME/QL]} Field Strength, $B$ [Gauss] : Total magnetic field
        strength.
  \item [\scriptsize$\blacksquare$] \textbf{[ME/QL]} Azimuth, $\chi$ [degrees] : Azimuthal angle of the
        transverse component (perpendicular to the LOS) of the vector magnetic field.  The azimuth has not
        yet been disambiguated at Level-1, so lies in the range [-90,+90]\degree.  The zero-direction
        points horizontally to the right, in the image plane.
  \item [\scriptsize$\blacksquare$] \textbf{[ME/QL]} Inclination, $\gamma$ [degrees] : The inclination of the
        magnetic field vector relative to the LOS.  Inclination lies in the range [0,180]\degree.  Quiet-sun
        values may tend to 90\degree\ because of noise in the polarization profiles (which have different
        sensitivity to the magnetic field) being misinterpreted as signal.  This does \textit{not} mean the
        quiet-Sun fields are truly horizontal!
  \item [\scriptsize$\blacksquare$] \textbf{[ME/QL]} Continuum Intensity, $I_{c}$ [counts] : The continuum
        intensity, calculated as the average Stokes \textit{I} signal in the continuum window.
  \item [\scriptsize$\blacksquare$] \textbf{[ME/QL]} Magnetic Filling-Factor, $\alpha$ [N/A] : The magnetic
        filling factor, defined as the fraction of the spatial pixel occupied by magnetic field.  The filling
        factor lies in the range [0,1].  This parameter is not inverted, but held fixed at the QL value for the
        duration of the inversion.
  \item [\scriptsize$\blacksquare$] \textbf{[ME]} Doppler Width, $\Delta\lambda_{D}$ [m\AA] : The Doppler width
        of the line.
  \item [\scriptsize$\blacksquare$] \textbf{[ME]} Line-to-Continuum Opacity Ratio, $\eta_{0}$ [N/A] : The
        ratio of absorption coefficient in the line to that of the continuum.
  \item [\scriptsize$\blacksquare$] \textbf{[ME]} Standard Error in Azimuth, $\sigma_{\chi}$ [degrees] : The
        standard uncertainty in the value of the azimuthal angle, $\chi$.
  \item [\scriptsize$\blacksquare$] \textbf{[ME]} Standard Error in Doppler Width, $\sigma_{\Delta\lambda_{D}}$ [m\AA] :
        The standard uncertainty in the value of the Doppler width, $\Delta\lambda_{D}$.
  \item [\scriptsize$\blacksquare$] \textbf{[ME/QL]} Degree of Polarization, $d_{\mathrm{pol}}$ [N/A] : The degree
        of polarization, defined as
        \begin{equation}
        d_{\mathrm{pol}} = \frac{\sqrt{\mathrm{\mathbf{max}}\left(Q_{\lambda}^{2}\right) +
                                       \mathrm{\mathbf{max}}\left(U_{\lambda}^{2}\right) +
                                       \mathrm{\mathbf{max}}\left(V_{\lambda}^{2}\right)}}{I_{c}}.
        \end{equation}
  \item [\scriptsize$\blacksquare$] \textbf{[ME]} Damping Parameter, $a_{\mathrm{dc}}$ [units of $\Delta\lambda_{D}$] : The atomic
        damping parameter of the line.
  \item [\scriptsize$\blacksquare$] \textbf{[ME]} Final $\chi^{2}$ Value, $\chi^{2}_{\mathrm{min}}$ [N/A] : The
        final returned value of the $\chi^{2}$ goodness-of-fit metric.
\end{itemize}

VFISV writes the output ME and QL FITS files with only a minimum FITS header.  The \solisvsm 6302v pipeline
later overwrites this with a more detailed header containing standard required FITS header keywords,
instrumental information from the observation metadata, and processing version information.  An example of
this \solisvsm Level-1 standard header is shown below, for the observation taken on 08 July 2015
(\texttt{k4v9s150708t175025\_oid114363777241750\_cleaned}) which resulted in the Level-1 file
\texttt{k4v91150708t175025.fts}.

\begin{verbatim}
SIMPLE  =                    T / file does conform to FITS standard
BITPIX  =                  -32 / number of bits per data pixel
NAXIS   =                    3 / number of data axes
NAXIS1  =                 2048 / length of data axis 1
NAXIS2  =                 2048 / length of data axis 2
NAXIS3  =                   12 / length of data axis 3
EXTEND  =                    T / FITS dataset may contain extensions
METAKEY0= '------------------- Standard Keywords ------------------'
DATE-OBS= '2015-07-08T17:50:25' / Observation date & start time (UTC)
DATE    = '2015-07-08T18:40:27' / file creation date (YYYY-MM-DDThh:mm:ss UT)
TELESCOP= 'solis   '           / Telescope
INSTRUME= 'vsm     '           / SOLIS instrument
OBSERVER= 'Detrick B.'         / Observer name
OBS-SITE= 'NSO/Tucson'         / Observation location
OBS-MODE= '6302v   '           / Observation mode
OBS-TYPE= 'full-scan'          / Observation type (dark, flat, full-scan)
STARTIME=           1436377825 / Obs-start in seconds since Jan. 1, 1970
STOPTIME=           1436379136 / Obs-stop in seconds since Jan. 1, 1970
QUALEND =                    0 / Completeness: 0- complete, 1- partial obs.
QUALSEE =                    0 / Seeing: 0- unknown, 1- bad, 5- excellent
QUALOBS =                    0 / Obs quality: 0- nominal, 1- generic bad
QUALCHK =                    0 / Qual-check method: 0- automatic
QMAXLEV =                    3 / Maximum processing level: Level-3 (Nominal)
IMGUNT01= '(Mx/cm^2)'          / Image units
IMGTYP01= 'Field Strength'     / Image type description
IMGUNT02= 'degrees '           / Image units
IMGTYP02= 'Azimuth '           / Image type description
IMGUNT03= 'degrees '           / Image units
IMGTYP03= 'Inclination'        / Image type description
IMGUNT04= 'Counts  '           / Image units
IMGTYP04= '630 nm Continuum Intensity' / Image type description
IMGUNT05= 'N/A     '           / Image units
IMGTYP05= 'Fill Factor'        / Image type description
IMGUNT06= 'milliAngstroms'     / Image units
IMGTYP06= 'Doppler Width'      / Image type description
IMGUNT07= 'N/A     '           / Image units
IMGTYP07= 'Line to cont opacity ratio' / Image type description
IMGUNT08= 'deg     '           / Image units
IMGTYP08= 'Std-error (Azimuth)' / Image type description
IMGUNT09= 'milliAngstroms'     / Image units
IMGTYP09= 'Std-error (dopw)'   / Image type description
IMGUNT10= 'N/A     '           / Image units
IMGTYP10= 'Percent Polarization' / Image type description
IMGUNT11= 'N/A     '           / Image units
IMGTYP11= 'Damping parameter'  / Image type description
IMGTYP12= 'Chi-square'         / Image type description
IMGUNT12= 'N/A     '           / Image units
PUBLISH =                    1 / Publish flag
METAKEY2= '------------------- Instrument Keywords ----------------'
OBS-ID  = 'oid114363777241750' / SOLIS obs ID
ENDSTATE= 'NOMINAL '           / Last DHS state
CAMTYPE = 'sarnoff '           / CCD camera type
NFRAMES =                  192 / Total number of frames per scanline
NSCANS  =                 2048 / Number of scanlines of observation
SCANINIT=                    0 / Initial scanline offset
BITSHIFT=                    2 / DAS Bit offset
GAPCOL1 =                  951 / First column within gap (one-indexed)
GAPCOL2 =                 1033 / Last column within gap (one-indexed)
M1-STAT =                    1 / Window cover position
M2-STAT =                15450 / M2 focus position
M3-STAT =                45759 / Calibration position
M6-STAT =               -13523 / Modulator position
M7-STAT =                -3740 / Littrow lens position
M9-STAT =                10247 / Beam-splitter position
GR-TILT =                 6266 / Grating tilt position
FOCALMSK=                    0 / Focal plane mask: out(0) or in(1)
WAVELNTH=                630.2 / Observed wavelength (nm)
METAKEY3= '------------------- Data Provenance --------------------'
PROVER0 =              13.1001 / Level-0 processing pipeline version
DDARKVER=               4.0831 / Dark-change cal-image version
VECTCAL = 'veccal_20150630.fts' / vector calibration used
DARK-ID = 'oid114363774301744' / Dark obsid
FLAT-ID = 'oid114363775501746' / Flat obsid
CHECKSUM= 'jGAjjE4jjE9jjE9j'   / HDU checksum updated 2015-07-08T18:40:28
DATASUM = '2662345718'         / data unit checksum updated 2015-07-08T18:40:28
END
\end{verbatim}

%

\bibliographystyle{te}
\bibliography{TechReport_NSO_NISP-2017-02}{}

\end{document}